\documentclass[11pt]{article}
\usepackage{graphicx}
\usepackage{amsmath}
\usepackage{epsf}
\usepackage{epsfig}
\usepackage{amssymb}
\usepackage{epstopdf}
\usepackage{latexsym}
\usepackage{subfigure}
\begin{document}
\vspace{.1cm}
\begin{center}
\vspace{.5cm}
{\bf\Large Reducing Constraints in a Higher Dimensional Extension of the Randall and Sundrum Model.}\\[.3cm]
\vspace{0.5cm}
Paul R.~Archer$^{a,}$\footnote{p.archer@sussex.ac.uk}
and 
Stephan J.~Huber$^{a,}$\footnote{s.huber@sussex.ac.uk} \\ 
\vspace{0.5cm} {\em  
$^a$Department of Physics \& Astronomy, University of Sussex, Brighton
BN1 9QH, UK }\\[.2cm]

\end{center}
\noindent
\vspace{0.3cm}

\begin{abstract}
In order to investigate the phenomenological implications of warped spaces in more than five dimensions, we consider a $4+1+\delta$ dimensional extension to the Randall and Sundrum model in which the space is warped with respect to a single direction by the presence of an anisotropic bulk cosmological constant. The Einstein equations are solved, giving rise to a range of possible spaces in which the $\delta$ additional spaces are warped. Here we consider models in which the gauge fields are free to propagate into such spaces. After carrying out the Kaluza Klein (KK) decomposition of such fields it is found that the KK mass spectrum changes significantly depending on how the $\delta$ additional dimensions are warped. We proceed to compute the lower bound on the KK mass scale from electroweak observables for models with a bulk $SU(2)\times U(1)$ gauge symmetry and models with a bulk $SU(2)_R\times SU(2)_L\times U(1)$ gauge symmetry. It is found that in both cases the most favourable bounds are approximately $M_{KK}\gtrsim 2$ TeV, corresponding to a mass of the first gauge boson excitation of about $4-6$ TeV. Hence additional warped dimensions offer a new way of reducing the constraints on the KK scale.      
\end{abstract}

\section{Introduction}
The Randall and Sundrum (RS) model \cite{Randall:1999ee} has proved a popular extension to the standard model for a number of reasons. Firstly it offers a non supersymmetric resolution to the gauge hierarchy problem and secondly it provides a natural model with which to describe flavour physics \cite{Grossman:1999ra, Gherghetta:2000qt, Huber:2000ie, Huber:2003tu, Agashe:2004cp, Bauer:2009cf}. Over the past decade much work has been done on the phenomenological implications of this model, see \cite{Csaki:2005vy, Gherghetta:2006ha, Davoudiasl:2009cd} for reviews, although much of this work has focused on the original 5D model. In particular, where the gauge fields are free to propagate in the bulk, corrections to the electroweak (EW) observables can naively put a lower bound on the Kaluza Klein (KK) mass scale of $M_{KK}\gtrsim 11$ TeV, corresponding to a mass of the first gauge boson excitation of about $27$ TeV \cite{Csaki:2002gy}. This bound can be reduced to $M_{KK}\gtrsim 2-3$ TeV  by allowing the fermions to propagate in the bulk and localising them towards the UV brane and at the same time introducing either large brane kinetic terms \cite{Carena:2002dz} or a bulk  $SU(2)_R\times SU(2)_L\times U(1)$ gauge symmetry\cite{Agashe:2003zs}. Although it should be noted, including anarchic Yukawa couplings and variable fermion positions increases this bound to $M_{KK}\gtrsim 3.5$ \cite{Delaunay:2010dw}.   

In spite of this, before the RS model can be considered a full resolution to the hierarchy problem it requires a UV completion. There has been considerable work on providing a UV completion via string theory, particularly with regard to the AdS/CFT correspondence, for example \cite{ArkaniHamed:2000ds, PerezVictoria:2001pa, Reece:2010xj}, but this work neccessarily requires models of more than five dimensions. Although the RS model is often seen as the low energy effective theory of $\rm{AdS}_5\times S^5$, it is quite possible that the $D>5$ additional dimensions are also warped \cite{Bao:2005ni, Klebanov:2000hb}.

In this paper we take a `bottom up' approach to extending the RS model in to more than 5D and consider a $D$ dimensional space warped by an anisotropic bulk cosmological constant. The principal motivation for such a choice is that it is one of the simplest extensions to the RS model that allows the investigation of the phenomenological implications of warping the $D>5$ additional dimensions and testing the robustness of the 5D phenomenology. Work has already been done looking at the phenomenology of $\rm{AdS}_5\times\mathcal{M}^\delta$ \cite{Davoudiasl:2002wz, Davoudiasl:2008qm} as well as $\rm{AdS}_D$ \cite{McDonald:2009hf, McDonald:2009md, McGuirk:2007er} and for additional work on 6D RS models see for example \cite{Collins:2001ni, Multamaki:2002cq, Koley:2006bh, Guo:2008ia}. The work very much continues from \cite{Archer:2010hh} where it was demonstrated that if the additional spaces are increasingly (decreasingly) warped towards the IR brane then the couplings of KK gauge fields propagating in the bulk will be suppressed (enhanced) relative to the zero mode couplings. This is nothing but the volume of the extra dimensions scaling the KK modes differently to those of the zero mode but it can result in reducing (increasing) the EW constraints on the model. 

The outline of this paper is as follows. In section \ref{Model} we solve the Einstein equations for a $4+1+\delta$ dimensional space with an anisotropic bulk cosmological constant and find three classes of solutions, those in which the $\delta$ extra dimensions are increasingly warped towards the IR brane, those which are decreasingly warped and the particular case where the extra dimensions are not warped. In section \ref{BGF} we carry out the KK decomposition of the gauge fields propagating in these extra dimensions and find that when the warping of the extra dimensions is close to one, there arises a splitting in the KK mass spectrum, from the $\delta$ extra KK towers, with a spacing of the same order of magnitude as that of the KK mass scale. If on the other hand the additional warping is increasing towards the IR brane then the spacing of the modes becomes very small and one gets a fine spacing in the KK spectrum. Conversely if the additional warping is decreasing, the spacing becomes large and the extra modes decouple from the low energy theory. We also compute the equations of motion for the gauge scalars and find them significantly different from conventional scalars. In section \ref{EWconst} we compute the lower bound on the KK mass scale from LEP 1 EW observables. We compute them for two models, one with a bulk $SU(2)\times U(1)$ gauge symmetry and the other with bulk $SU(2)_R\times SU(2)_L\times U(1)$ gauge symmetry. In both models we localise the fermions to the IR or UV brane\footnote{Here fermions localised on the UV brane are an approximation of a model with bulk fermions localised towards the UV brane.}. As expected the EW constraints are significantly reduced in spaces which are increasingly warped towards the IR brane although it is not possible in the spaces studied here to sizeably reduce the constraint from that of 5D models with a bulk custodial symmetry and fermions localised towards the UV brane. Hence we conclude that the most favorable bound for the KK mass scale is roughly $M_{KK}\gtrsim 2$ TeV.  Nevertheless, this mechanism offers a new way of reducing the constraint on the KK scale and one in which the fermions can still be localised on the IR brane.     

\section{The Model}\label{Model}
Here we consider a relatively simple extension of the Randall and Sundrum Model. It concerns the classical solutions of a $4+1+\delta$ dimensional space, bounded by two 3 branes, described by  
\begin{eqnarray}
S=\int d^{5+\delta}x\sqrt{G}\left [\Lambda-\frac{1}{2}M^{3+\delta}R+\mathcal{L}_{\rm{bulk}}\right]+\int d^4x\sqrt{g_{\rm{ir}}}\left [\mathcal{L}_{\rm{ir}}+V_{\rm{ir}}\right ] \nonumber \\
 +\int d^4x\sqrt{g_{\rm{uv}}}\left [\mathcal{L}_{\rm{uv}}+V_{\rm{uv}}\right ]\label{action},
\end{eqnarray}   
where $g^{\rm{ir/uv}}_{\mu\nu}$ is the induced metric $G_{MN}\delta^M_\mu\delta^N_\nu\delta(r-r_{\rm{ir/uv}})\delta^{(\delta)}(\theta-\theta_{\rm{ir/uv}})$. The co-ordinates run over $(x_\mu,r,\theta_1\dots\theta_\delta)$, we also collectively write $\theta=(\theta_1\dots\theta_\delta)$. We allow for an anisotropic bulk cosmological constant of the form
\begin{equation}\label{lambda}
\Lambda=\begin{pmatrix}
\Lambda \eta_{\mu\nu } &&&& \\
&\Lambda_5 &&& \\
&&\Lambda_\theta &&\\
&&&\ddots &\\
&&&&\Lambda_\theta\\
\end{pmatrix}.
\end{equation}
It should be stressed that this is very much a toy model. If, for example, we were to consider such a cosmological constant arising from an $r$, $\theta_1$ dependent VEV of some bulk scalar in 6D
\begin{displaymath}
S=\int d^6x\sqrt{-G}\left [-\frac{1}{2}M^4R+\Lambda+\frac{1}{2}|\partial_M\Phi|^2-V(\Phi)\right ],
\end{displaymath}
the corresponding energy-momentum tensor would be
\begin{equation*}
T_{MN}=\begin{pmatrix}
\left [-\Lambda-\frac{1}{2}\Phi^{\prime\;2}-\frac{1}{2}\dot{\Phi}^2+V(\Phi)\right ]\eta_{\mu\nu } && \\
&-\Lambda+\frac{1}{2}\Phi^{\prime\;2}-\frac{1}{2}\dot{\Phi}^2+V(\Phi)& \\
&&-\Lambda-\frac{1}{2}\Phi^{\prime\;2}+\frac{1}{2}\dot{\Phi}^2+V(\Phi)\\
\end{pmatrix}.
\end{equation*}
Here $^\prime$ and $\dot{}$ denotes derivative with respect to (w.r.t) $r$ and $\theta_1$. Hence (\ref{lambda}) could be realised when the derivatives dominate over the potential and the VEV is linearly dependent on $r$ and $\theta_1$. Arguably any realistic approach should allow for a energy-momentum tensor dependent on $r$ and $\theta$. However this would lead to warping in multiple directions, complicated wavefunctions and an enlarged parameter space. Although here we consider very much a bottom up approach, it is of course partially motivated by the AdS throats that arise from the AdS/CFT correspondence. Hence we use a metric ansatz warped w.r.t a single `preferred' direction $r$,         
\begin{equation}
ds^2=e^{-2A(r)}\eta_{\mu\nu}dx^\mu dx^\nu-dr^2-\sum_{i=1}^{\delta}e^{-2C(r)}d\theta_i^2.
\end{equation}
Here the space runs over the intervals $r\in [0,R]$ and $\theta_i\in [0,R_\theta]$. Throughout this paper $M,N$ run over all the space time indices $0,\dots,5+\delta$, while $\mu,\nu$ run from $0,\dots, 3$ and $i,j,k$ run over the additional dimensions $1,\dots,\delta $. For simplicity we take the internal geometry to be toroidal. Neglecting the contribution from any bulk fields the Einstein equations then follow from (\ref{action})
\begin{eqnarray}
6A^{\prime\;2}-3A^{\prime\prime}+3\delta A^{\prime}C^{\prime}-\delta C^{\prime\prime}+\frac{(\delta+1)!}{2!(\delta-1)!}C^{\prime\;2}=-\frac{1}{M^{3+\delta}}\Big [\Lambda+\hspace{1.5cm}\nonumber\\+V_{ir}e^{\delta C}\delta(r-r_{ir})\delta^{(\delta)}(\theta-\theta_{ir})+V_{uv}e^{\delta C}\delta(r-r_{uv})\delta^{(\delta)}(\theta-\theta_{uv})\Big ]\nonumber\\
6A^{\prime\;2}+\frac{\delta !}{2!(\delta-2)!}C^{\prime\; 2}+4\delta A^{\prime}C^{\prime}=-\frac{\Lambda_5}{M^{3+\delta}}\nonumber\\
4(\delta-1)A^{\prime}C^{\prime}-(\delta-1)C^{\prime\prime}+\frac{\delta !}{2!(\delta-2)!}C^{\prime\; 2}+10A^{\prime\;2}-4A^{\prime\prime}=-\frac{\Lambda_\theta}{M^{3+\delta}}.\label{einstein}
\end{eqnarray}

\begin{figure}[!t]
\begin{center}
\includegraphics[width=5.5in]{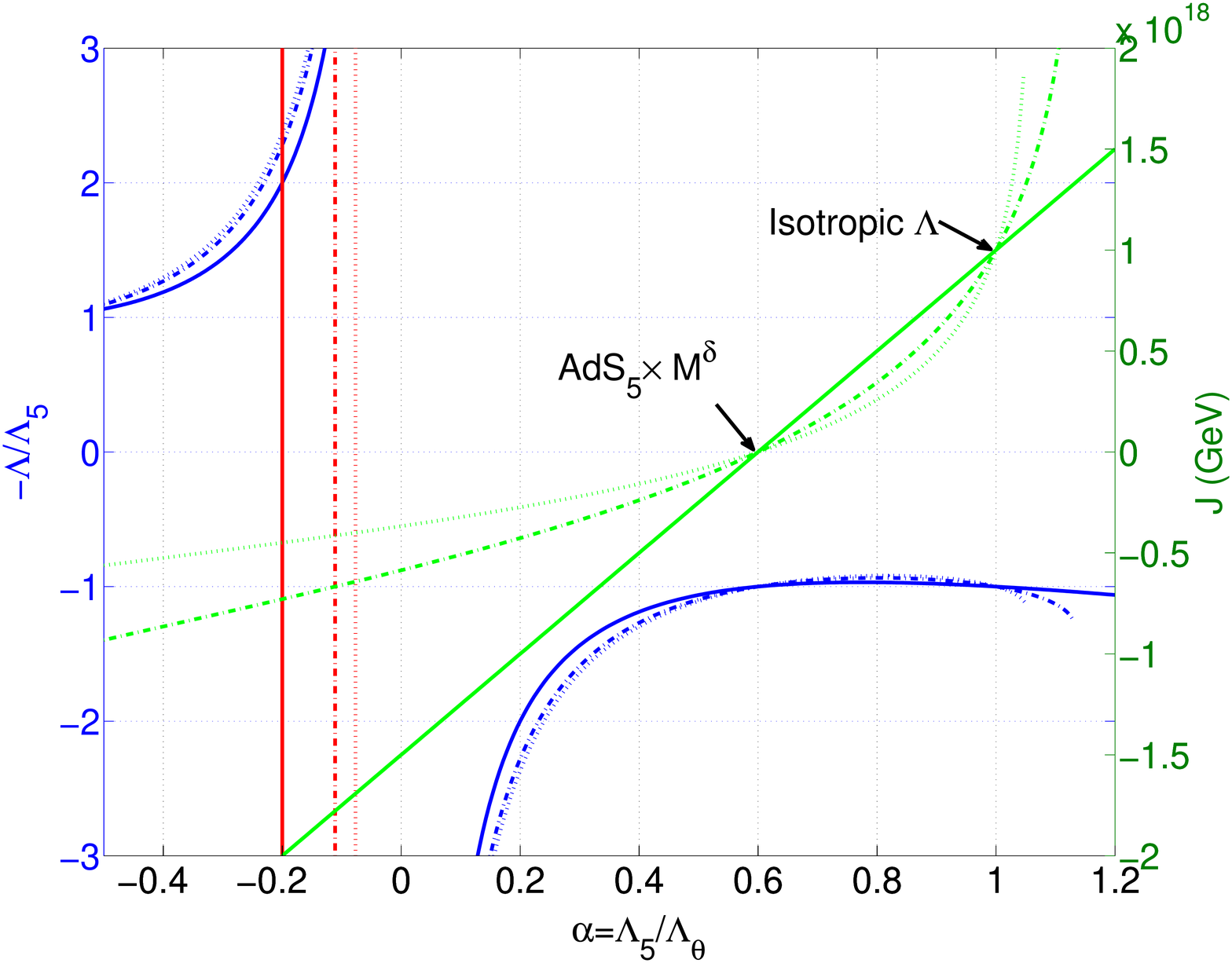}
\caption{The solutions to the Einstein equations (\ref{einstein}) for 6D (solid lines), 8D(dash-dot lines) and 10D (dot-dot lines). $J$ is plotted in green and $-\frac{\Lambda}{\Lambda_5}$ in blue. The red lines correspond to $2k+\delta J=0$ and hence the spaces to the left of them will have exponentially suppressed fundamental Planck masses. Here we have fixed $\Omega\equiv e^{kR}=10^{15}$ and $M_{KK}\equiv \frac{k}{\Omega}=1$ TeV.}
\label{JLamb}
\end{center}
\end{figure}

\begin{figure}[!t]
\begin{center}
\includegraphics[width=6in]{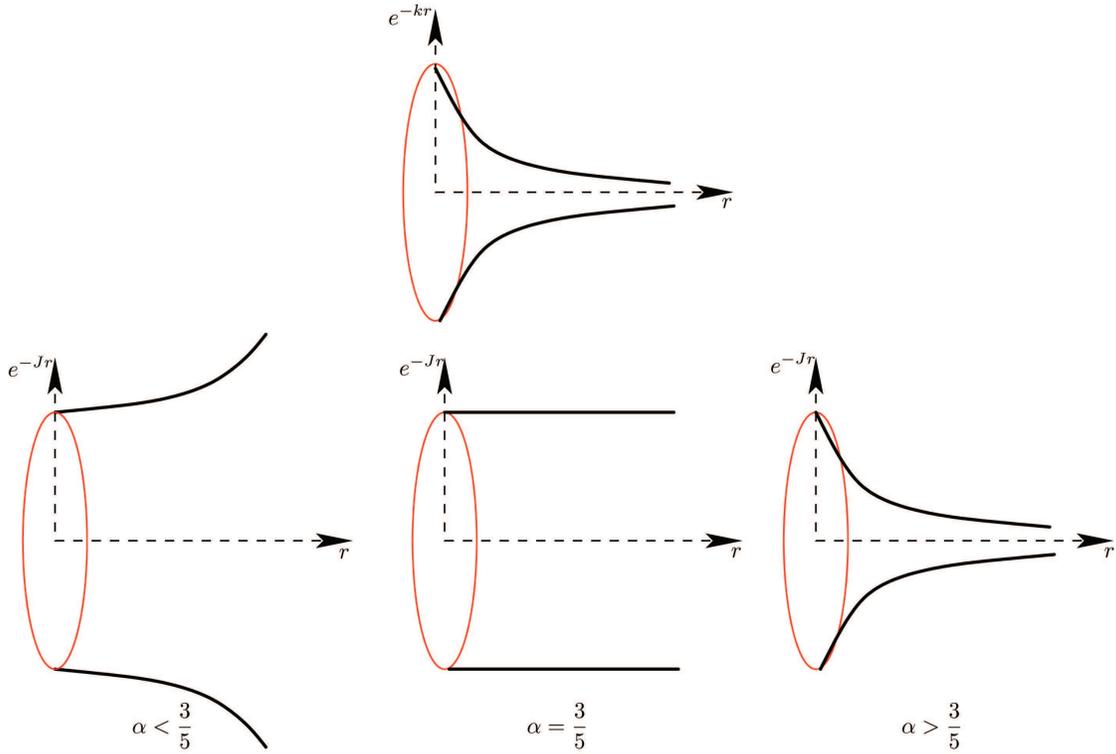}
\caption{The warping of the radii of the additional dimensions.}
\label{throats}
\end{center}
\end{figure}

Only two of these equations are independent and hence admit the solution (although not unique solution)
\begin{equation*}
A(r)=kr\qquad \mbox{and} \qquad C(r)=Jr,
\end{equation*}
where $k$ and $J$ are constants, depending on $\Lambda$, $\Lambda_5$ and $\Lambda_\theta$. Here we introduce the parameter
\begin{equation*}
\alpha\equiv\frac{\Lambda_5}{\Lambda_\theta}.
\end{equation*}
Then in 6D, as found in \cite{Kogan:2001yr}, (\ref{einstein}) leads to
\begin{displaymath}
k=\pm\sqrt{-\frac{\Lambda_5}{10\alpha M^4}},\qquad J=\frac{5}{2}\alpha k-\frac{3}{2}k,\qquad \Lambda=\left (\frac{3}{8\alpha}+\frac{5}{8}\alpha\right )\Lambda_5.
\end{displaymath}
See also \cite{Multamaki:2002cq} for 6D work in this direction. In more than six dimensions the equivalent equations have a more complicated form and are plotted in figure \ref{JLamb}. There are essentially three classes of spaces as illustrated in figure \ref{throats}. When $\alpha<\frac{3}{5}$, the radii of the $\delta$ additional dimensions increases as one moves towards the IR brane and according to \cite{Archer:2010hh}, one would anticipate the KK modes of bulk gauge field having reduced couplings to matter on the IR brane. Conversely, where $\alpha>\frac{3}{5}$, the additional dimensions are shrinking towards the IR brane and one would anticipate enhanced couplings. Note that an isotropic cosmological constant ($\Lambda=\mbox{diag}(\Lambda,\Lambda,\dots,\Lambda)$) would belong to this class of solutions and would naively have large EW corrections. The third solution corresponding to $\alpha=\frac{3}{5}$ (or $\Lambda= \mbox{diag}(\Lambda\eta_{\mu\nu},\frac{3}{5}\Lambda,\Lambda,\dots,\Lambda)$) would correspond to the $\rm{AdS}_5\times M^{\delta}$ space studied in \cite{Davoudiasl:2002wz, Davoudiasl:2008qm}. Note also that spaces with $\alpha<0$ are valid but would correspond to a partially deSitter space with $\Lambda_5>0$.  

In (\ref{action}) the space was cut off by two 3 branes. However, in constructing a realistic model with branes in, one faces a number of conflicting problems. Firstly it has been known for some time \cite{PhysRevD.23.852} that matching the solutions across a brane, with a tension, of codimension two or higher typically gives rise to a singularity at the location of the brane. There are known solutions that avoid this singularity particularly in six dimensions \cite{Bayntun:2009im}, for example, including a bulk Gauss-Bonnet term \cite{Bostock:2003cv}.  However this problem becomes increasingly severe for branes of codimension three and higher \cite{Gherghetta:2000jf, Charmousis:2001hg}. Here we are primarily interested in the low energy phenomenological implications of these geometries, particularly with regard to electroweak constraints. This gives rise to two additional problems. Firstly reproducing a four dimensional effective chiral theory or in other words finding a suitable orbifold such that the zero mode of the higher dimensional fermion is a 2 component Weyl spinor. Again this is possible in six dimensions \cite{Dobrescu:2004zi} but becomes increasingly difficult in higher dimensional models. Secondly, as explained in the appendix, localising the Higgs on a brane essentially mixes the KK modes of the W and Z and gives rise to tree level EW constraints. When the Higgs is localised on a codimension two brane, or higher, the EW constraints will involve a sum over multiple KK towers, which is typically divergent. Hence the EW constraints are only perturbatively computable when the Higgs is localised on a codimension one brane. 
     
It should be stressed that these are problems that all $D>5$ dimensional models face. Clearly it is beyond the scope of this paper to resolve all these issues for all dimensions. One possibility would be to restrict this study to a six dimensional space bounded by two 4 branes, in which many of these problems are simpler to solve. However the central results of this paper are related to the scaling, by the geometry, of the couplings and masses of the KK particles and it would be useful to know how these results change with dimensionality. Hence here we will consider a toy model in which we neglect the back reaction of both the bulk fields and brane tensions on the resulting geometry. We will also, for simplicity, localise the fermions to 3 branes and the Higgs on a codimension one brane.     

In order to resolve the hierarchy problem we require that the fundamental Higgs mass and the curvature, $k$ and $J$, are of the same order of magnitude as that of the fundamental Planck mass. If the Higgs is localised on the IR brane then the effective mass is suppressed by gravitational red shifting,
\begin{equation*}
m_{\rm{eff}}=e^{-kR}m_{\rm{fund}}\equiv\Omega^{-1}m_{\rm{fund}},
\end{equation*}
while the Planck mass is
\begin{equation*}
M_P^2=M^{3+\delta}\frac{(1-e^{-(2k+\delta J)R})(R_{\theta})^\delta}{(2k+\delta J)}.
\end{equation*}
Hence when $2k+\delta J>0$, as in the 5D RS model, the volume effects are of order one and a warpfactor $\Omega\sim 10^{15}$ is required to resolve the hierarchy problem (assuming $R_{\theta}^{-1}\sim M \sim M_P$). Conversely if $2k+\delta J<0$ then one moves towards a `warped ADD' scenario in which the Planck mass is exponentially suppressed by the volume of the additional dimensions, hence introducing an additional hierarchy between the AdS curvature and the Planck mass. Also recent work has indicated that there may be problems with unitarity in lowering the Planck scale to about a TeV \cite{Atkins:2010re} and hence here we focus on spaces with $2k+\delta J>0$. Of course it is always possible that the fundamental scale is neither the Planck scale nor the EW scale but some intermediate scale. In other words one could, for example, increase the size of $R_\theta$ and hence the required warp factor would be reduced. If this was the case the basic results of this paper would not change significantly although the numbers of course would. Having said that throughout this paper we fix the warp factor to be $\Omega=10^{15}$. It is this that results in the infinity at $\alpha=0$ in figure \ref{JLamb}. 

\section{Bulk Gauge Fields}\label{BGF}     
In this section we consider the KK decomposition of a gauge field propagating in the spaces described in section \ref{Model}
\begin{equation}   
ds^2=e^{-2kr}\eta_{\mu\nu}dx^\mu dx^\nu-dr^2-\sum_{i=1}^{\delta}e^{-2Jr}d\theta_i^2.
\end{equation} 
Let us consider an abelian bulk gauge field with an action 
\begin{equation*}
S=\int d^{5+\delta}x\sqrt{-G}\left (-\frac{1}{4}F_{MN}F^{MN}\right ),
\end{equation*} 
where $F_{MN}=\partial_M A_N-\partial_N A_M$. Working in the $R_\xi$ gauge where the gauge fixing term is chosen in order to cancel the $\partial_\mu A^\mu\partial_{i,5}A^{i,5}$ terms
\begin{equation*}
S_{\rm{GF}}=-\int d^{5+\delta}x\,\frac{1}{2\xi}e^{-\delta Jr}\left (\partial_\mu A^{\mu}-\xi e^{\delta Jr}\left (\partial_5(e^{-(2k+\delta J)r}A_5)+\sum_{i=1}^{\delta}e^{-(2k+\delta J-2J)r}\partial_iA_i\right )\right )^2.
\end{equation*}
Varying the action yields the equations of motion
\begin{eqnarray}
-e^{-\delta Jr}\partial^\nu F_{\nu\mu}+\partial_5(e^{-(2k+\delta J)r}\partial_5A_\mu)+\sum_{i=1}^{\delta}e^{-(2k+\delta J-2J)r}\partial_i^2A_\mu-\frac{1}{\xi}e^{-\delta Jr}\partial_\mu(\partial_\nu A^\nu)=0,\label{AuEOM}\\
e^{-(2k+\delta J)r}\partial_\mu\partial^\mu A_5-e^{-(4k+\delta J-2J)r}\sum_{i=1}^{\delta}\partial_i(\partial_iA_5-\partial_5A_i)-\hspace{5cm}\nonumber\\
-\xi e^{-(2k+\delta J)r}\left (\partial_5(e^{\delta Jr}\partial_5(e^{-(2k+\delta J)r}A_5))+\sum_{i=1}^{\delta}\partial_5(e^{-(2k+2J)r}\partial_iA_i)\right )=0,\label{A5EOM}\\
e^{-(2k+\delta J-2J)r}\partial_\mu\partial^\mu A_i-\partial_5\left(e^{-(4k+\delta J-2J)r}(\partial_5A_i-\partial_iA_5)\right )-\sum_{j=1}^{\delta}e^{-(4k+\delta J-4J)r}\partial_j(\partial_jA_i-\partial_iA_j)-\nonumber\\
-\xi e^{-(2k+\delta J-2J)r}\left (e^{\delta Jr}\partial_i\partial_5(e^{-(2k+\delta J)r}A_5)+\sum_{j=1}^{\delta}e^{-(2k+2J)r}\partial_i\partial_jA_j\right )=0.\label{AiEOM}
\end{eqnarray}
Note that in computing the above equations we have assumed the boundary terms vanish i.e. that the $A_\mu$, $A_5$ and $A_i$ fields have either Neumann ($+$) or Dirichlet ($-$) boundary conditions at all boundaries. 

\subsection{The KK Decomposition of the Gauge Fields}
Let us first consider the 4D gauge fields, $A_\mu$. If we define their KK masses by $\partial^\nu F_{\nu\mu}+\frac{1}{\xi}\partial_\mu(\partial_\nu A^{\nu})=-m_n^2A_\mu$, then carrying out the usual KK decomposition
\begin{equation}\label{KKdecomp}
A_\mu=\sum_n A_\mu^{(n)}(x^\mu)f_n(r)\Theta_n(\theta_1,\dots,\theta_\delta),
\end{equation}
such that
\begin{equation}\label{gaugeOrthog}
\int drd^\delta\theta \;e^{-\delta Jr}f_nf_m\Theta_{n}\Theta_{m}=\delta_{nm}.
\end{equation}
If we let $\partial_i^2\Theta_n=-\frac{l_i^2}{R_\theta^2}\Theta_n$\footnote{Here we can allow for general $l_i$, as arising from the eigenvalues of the Laplacian of a generic internal geometry. Note however in (\ref{gaugeOrthog}) we have assumed toroidal additional dimensions and hence not included the correspondng weighting factor.}, where $l_i$ are constants dependent on the internal space, then (\ref{AuEOM}) takes the form;
\begin{equation}
f_n^{\prime\prime}-(2k+\delta J)f_n^{\prime}-\sum_{i=1}^\delta e^{2Jr}\frac{l_i^2}{R_\theta^2}f_n+e^{2kr}m_n^2f_n=0.\label{gaugeeqn}
\end{equation}
First note that there are now essentially $1+\delta$ KK towers where the mass spectrum depends on the possible $l_i$ values. We would like to know the masses of these KK modes but unfortunately (\ref{gaugeeqn}) has no analytical solution. To get a feel for the solutions of (\ref{gaugeeqn}) we make the substitution
\begin{equation}\label{Xdef}
x^2=\frac{e^{2kr}m_n^2-e^{2Jr}\sum_{i=1}^\delta\frac{l_i^2}{R_\theta^2}}{\gamma^2}\qquad\mbox{such that}\qquad x^{\prime}=\gamma x
\end{equation}
and hence
\begin{equation}\label{gammaeqn}
\gamma^\prime=\left(\frac{ke^{2kr}m_n^2-Je^{2Jr}\sum_{i=1}^\delta\frac{l_i^2}{R_\theta^2}}{e^{2kr}m_n^2-e^{2Jr}\sum_{i=1}^\delta\frac{l_i^2}{R_\theta^2}}\right )\gamma-\gamma^2.
\end{equation}
\begin{figure}[t]
\begin{center}
\begin{tabular}{cc}
\includegraphics[width=3.2in]{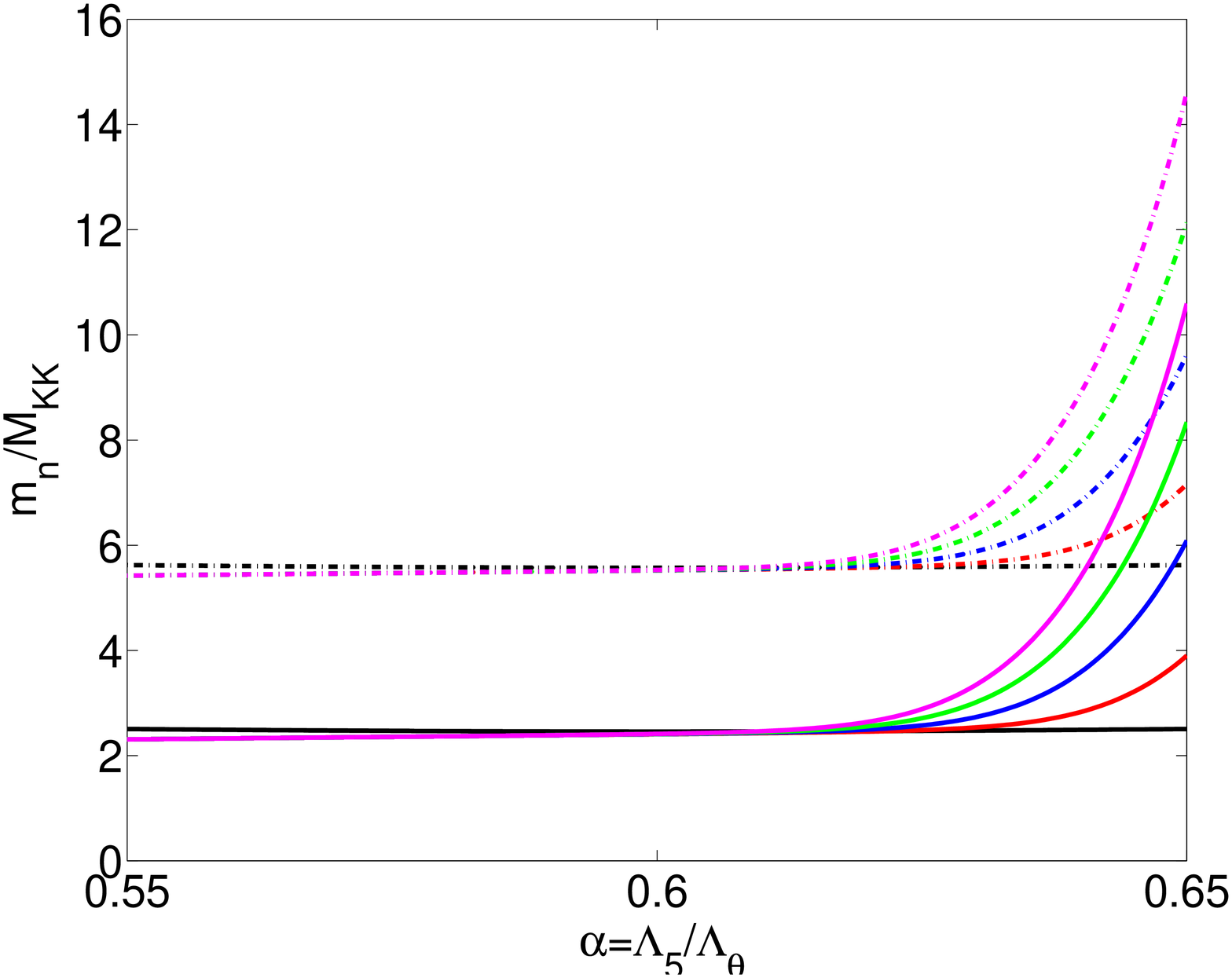}&
\includegraphics[width=3.2in]{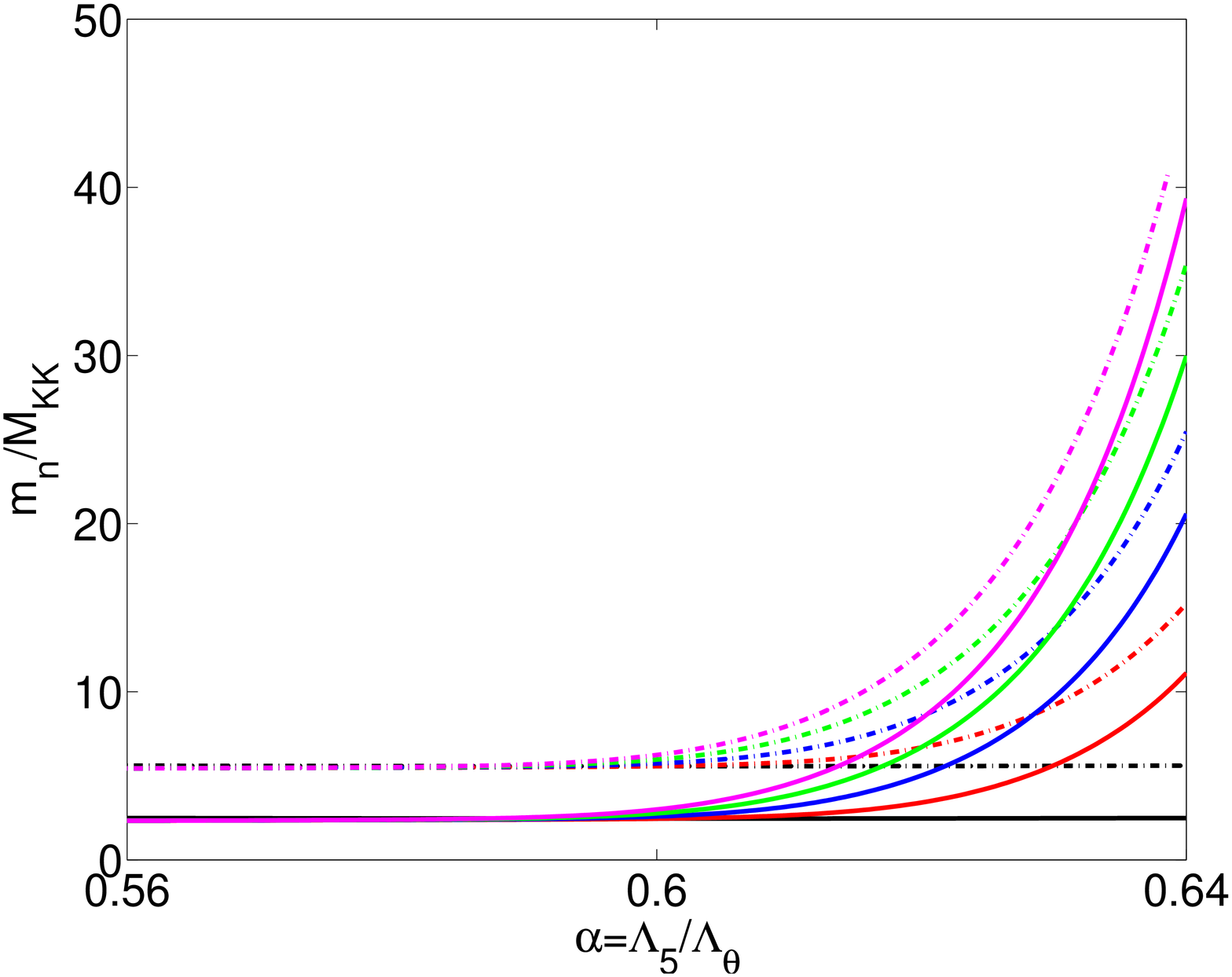}\\
\end{tabular}
\caption{The first two mass eigenvalues of gauge KK modes in six dimensions, for $l_1=0$ (black), $l_1=1$ (red), $l_1=2$ (blue), $l_1=3$ (green) and $l_1=4$ (purple).  Here $R_\theta=R$ on the left and $R_\theta=0.1R$ on the right. $\Omega\equiv e^{kR}=10^{15}$ and $M_{KK}\equiv\frac{k}{\Omega}=1$TeV. }
\label{gaugedegfig}
\end{center}
\end{figure}   
It is important to note that $\gamma$ is typically going to be of the same order of magnitude as $J$ and $k$. In the IR region, where the gauge fields are naturally localised, $\gamma \approx$ constant and (\ref{gaugeeqn}) can be approximated to 
\begin{equation*}
\ddot{f}+\left (1-\frac{2k+\delta J}{\gamma}\right )\frac{1}{x}\dot{f}+f\approx 0
\end{equation*}
where $\dot{}$ denotes derivative w.r.t. $x$. This can now be solved to give
\begin{equation}
f(x)\approx Nx^{\frac{1}{2}\frac{2k+\delta J}{\gamma}}\left (\mathbf{J}_{-\frac{1}{2}\frac{2k+\delta J}{\gamma}}(x)+\beta \mathbf{Y}_{-\frac{1}{2}\frac{2k+\delta J}{\gamma}}(x) \right ).
\end{equation}
The mass eigenvalues will then be largely determined by the zeros of the Bessel functions and hence, assuming $\frac{1}{2}\frac{2k+\delta J}{\gamma}\sim \mathcal{O}(1)$, the KK masses would roughly follow from (\ref{Xdef}) to be
\begin{equation}\label{MNapprox}
m_n\sim X_n\frac{\sqrt{\gamma^2+\frac{e^{2JR}}{R_\theta^2}\sum_i^{\delta}l_i^2}}{e^{kR}}.
\end{equation}
Here $X_n$ will be some $\mathcal{O}(1)$ number dependent on the roots of the Bessel functions, the boundary conditions and any UV effects. Clearly the spacing of KK modes will depend on the warping of the additional dimensions. If for example we take $R_\theta^{-1}\sim M_P$, $M_{KK}\equiv\frac{k}{\Omega}\sim 1$ TeV and $\Omega\sim10^{15}$ then:
\begin{itemize} 
\item If $J>0$ (i.e. $\alpha>\frac{3}{5}$) then $\frac{e^{JR}}{R_\theta}\gg\gamma\sim k$ and the $l_i\neq 0$ KK modes would gain masses far larger than $M_{KK}$ and essentially decouple from the low energy theory. Hence one would be left with just the KK tower corresponding to the $l_i=0$ modes.
\item If $J\sim 0$ then $\frac{e^{JR}}{R_\theta}\sim\gamma$ and the $l_i\neq 0$ KK modes would have masses of $\mathcal{O}(M_{KK})$. Hence the splitting in the KK spectrum would be apparent at the KK scale. For a specific example of the graviton KK spectrum in this case see \cite{Davoudiasl:2002wz, Davoudiasl:2008qm}.
\item If $J<0$ then the masses of the $l_i\neq 0$ modes would not shift much from those of the $l_i=0$ modes. Hence one would introduce a fine splitting in the KK spectrum. In an ideal world, the observation of such a splitting at the LHC would surely be a clear cut signal of extra dimensions and could even allow one to read of the dimensionality and topology of space-time.     
\end{itemize}

\begin{figure}[t]
\begin{center}
\begin{tabular}{cc}
\includegraphics[width=3.2in]{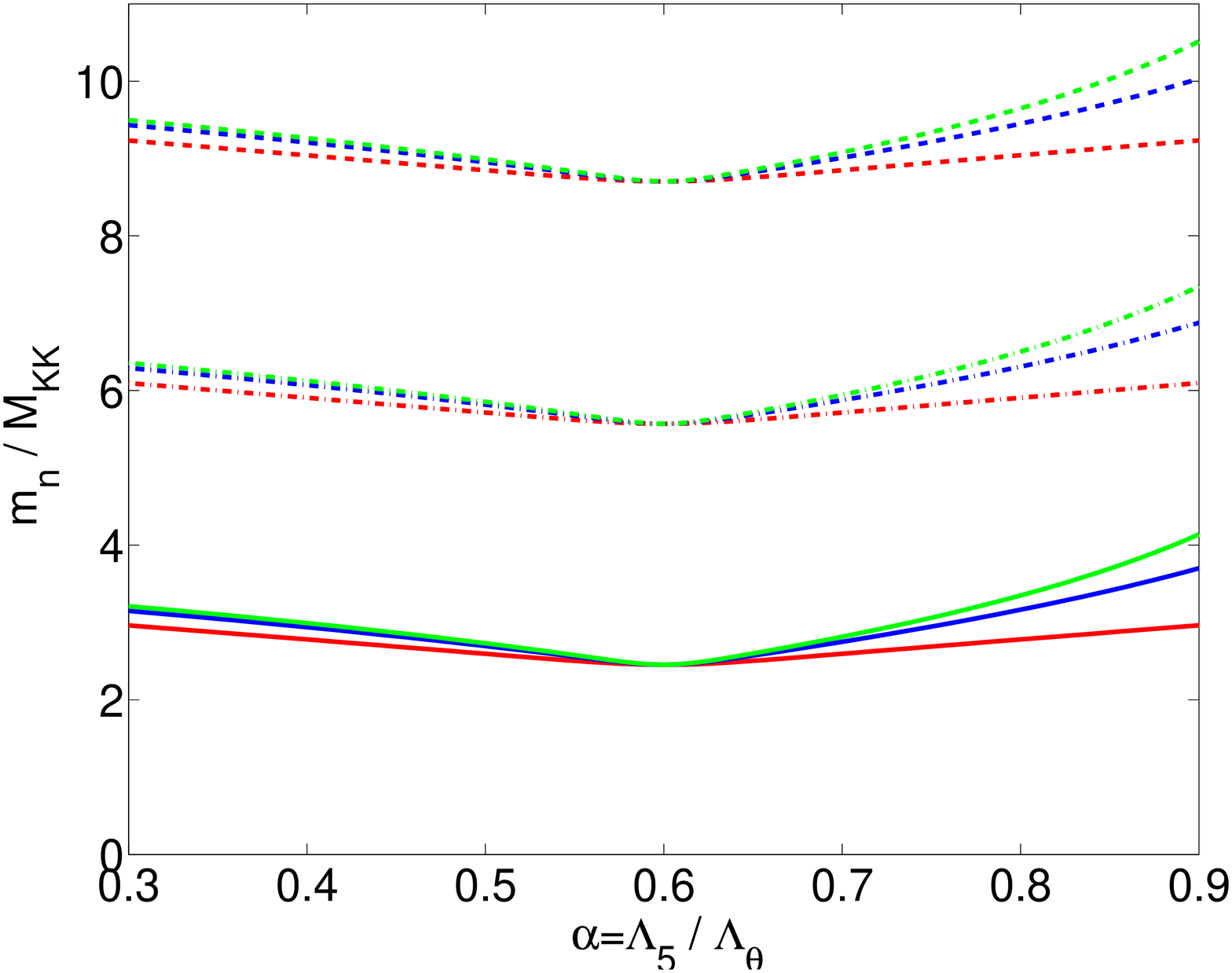}&
\includegraphics[width=3.2in]{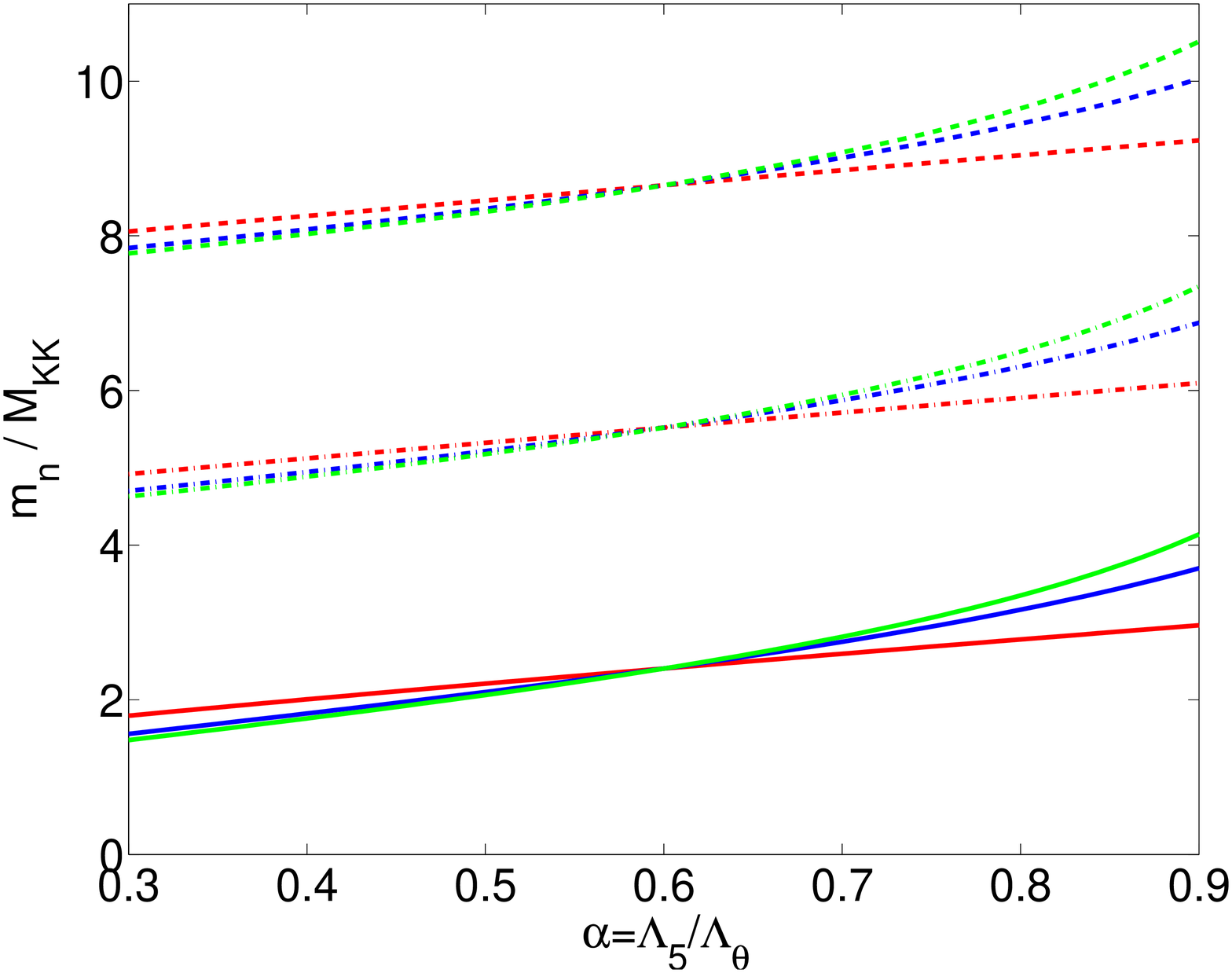}\\
\end{tabular}
\caption{First three mass eigenvalues of the $l_i=0$ gauge KK modes with the (IR,UV) boundary conditions $(++)$, on the left and $(-+)$ on the right. The gauge field propagates in 6D (red), 8D (blue) and 10D (green). Here $\Omega\equiv e^{kR}=10^{15}$ and $M_{KK}\equiv\frac{k}{\Omega}=1$TeV. }
\label{gaugemassesfig}
\end{center}
\end{figure}

One should be skeptical about the validity of neglecting the contribution from the UV region of the space to $\gamma$ and hence the KK scale. However (\ref{MNapprox}) is found to be in rough agreement with numerical solutions of (\ref{gaugeeqn}), plotted for six dimensions, in figure \ref{gaugedegfig}. It is also possible to analytically solve (\ref{gaugeeqn}) for the $l_i=0$ modes to get,
\begin{equation}\label{gaugewave}
f_n(r)=Ne^{vr}\left [\mathbf{J}_{-\frac{v}{k}}\left(\frac{m_ne^{kr}}{k}\right )+\beta \mathbf{Y}_{-\frac{v}{k}}\left (\frac{m_ne^{kr}}{k}\right )\right ],
\end{equation}
where $v=\frac{1}{2}(2k+\delta J)$ and $N$ is computed from (\ref{gaugeOrthog}). The mass eigenvalues are plotted in figure \ref{gaugemassesfig}. An important point for distinguishing between models at the LHC is that, as in the 5D RS model, in spaces with $J\geq0$ there is not a significant difference between the masses of fields with Neumann or Dirichlet boundary conditions on the IR brane. On the other hand where $J<0$ one starts reducing the root of the Bessel function and introducing a large mass difference between modes with different boundary conditions.

\begin{figure}[th!]
\begin{center}
\begin{tabular}{cc}
\includegraphics[width=3in]{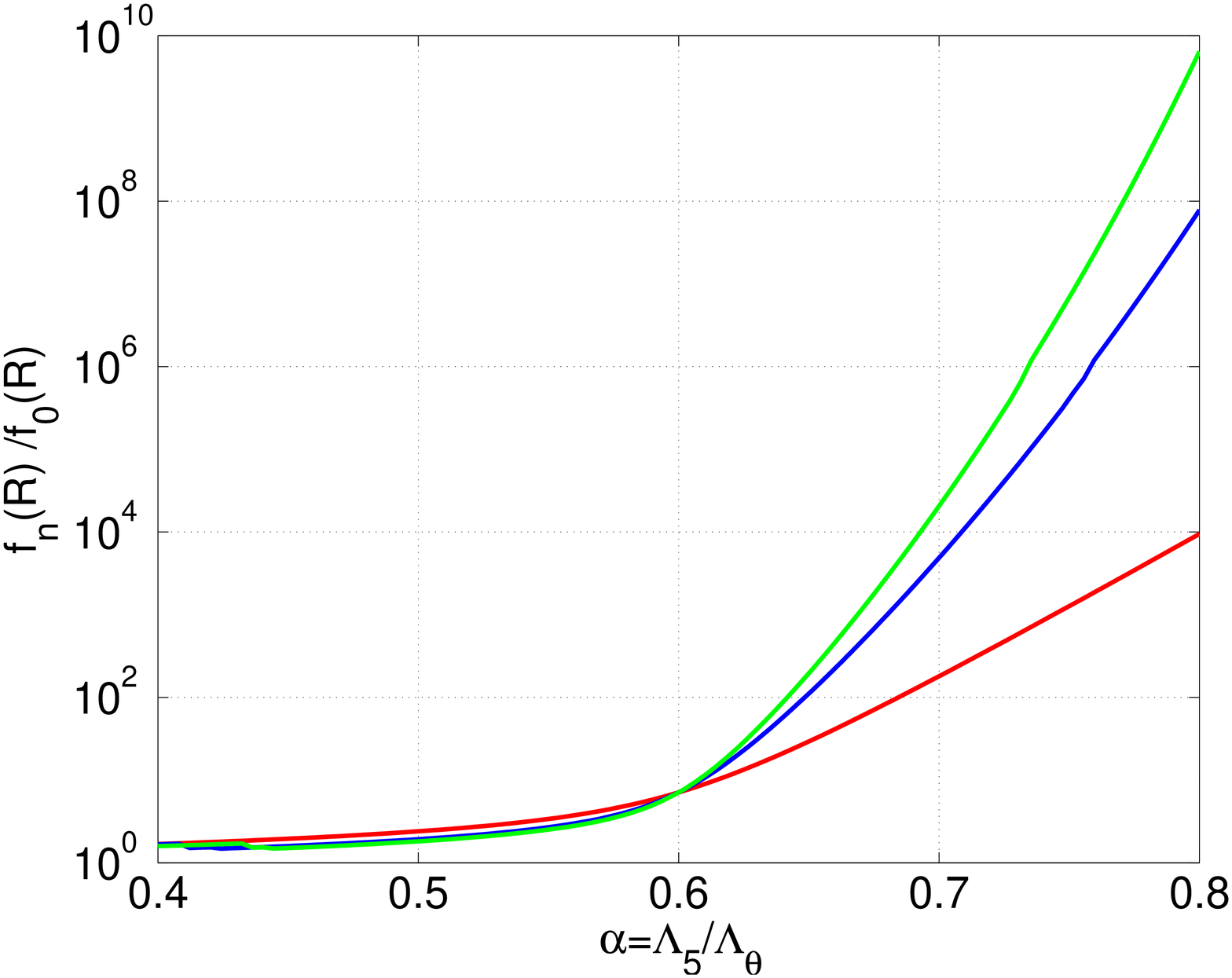}&
\includegraphics[width=3in]{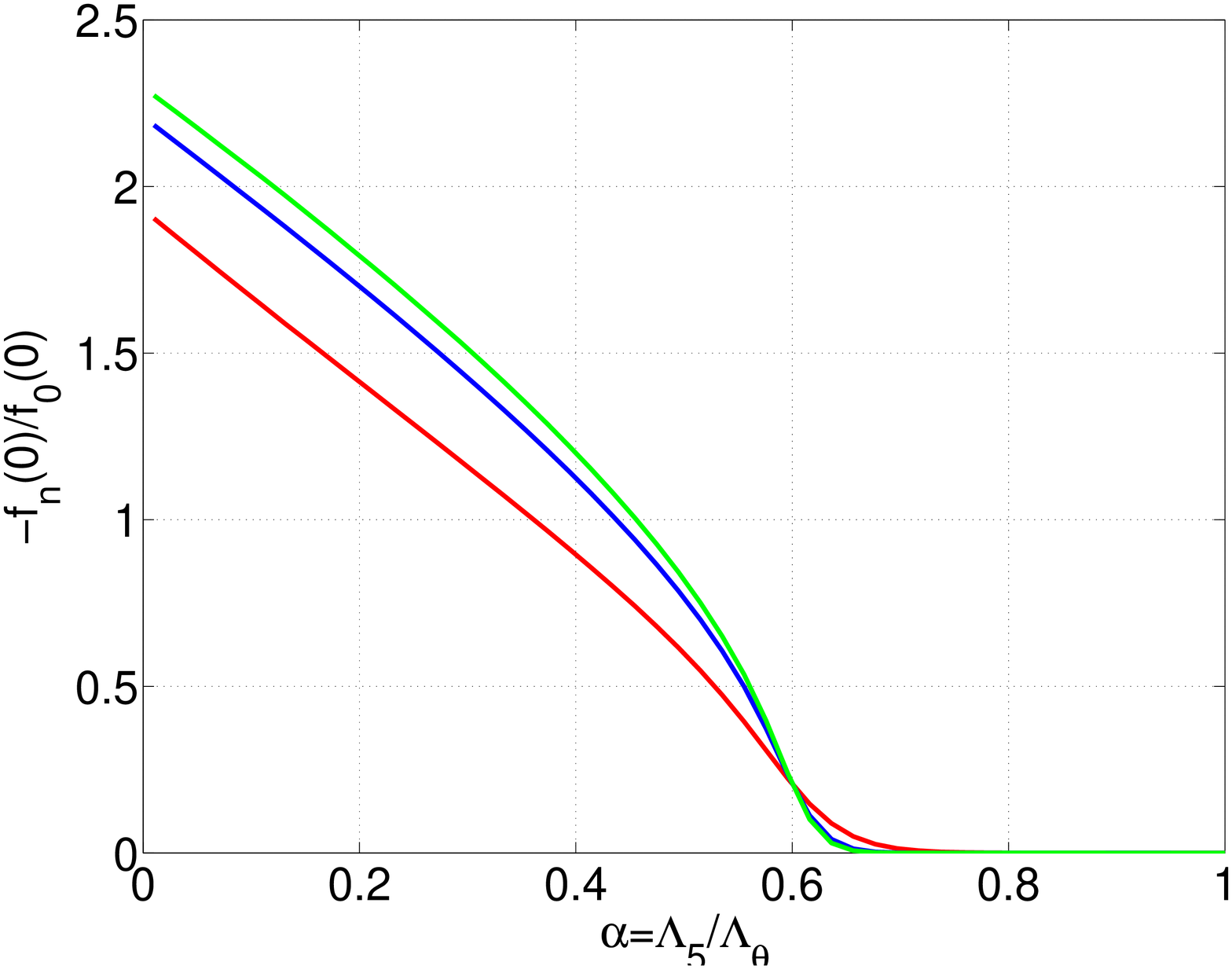}\\
\end{tabular}
\includegraphics[width=3in]{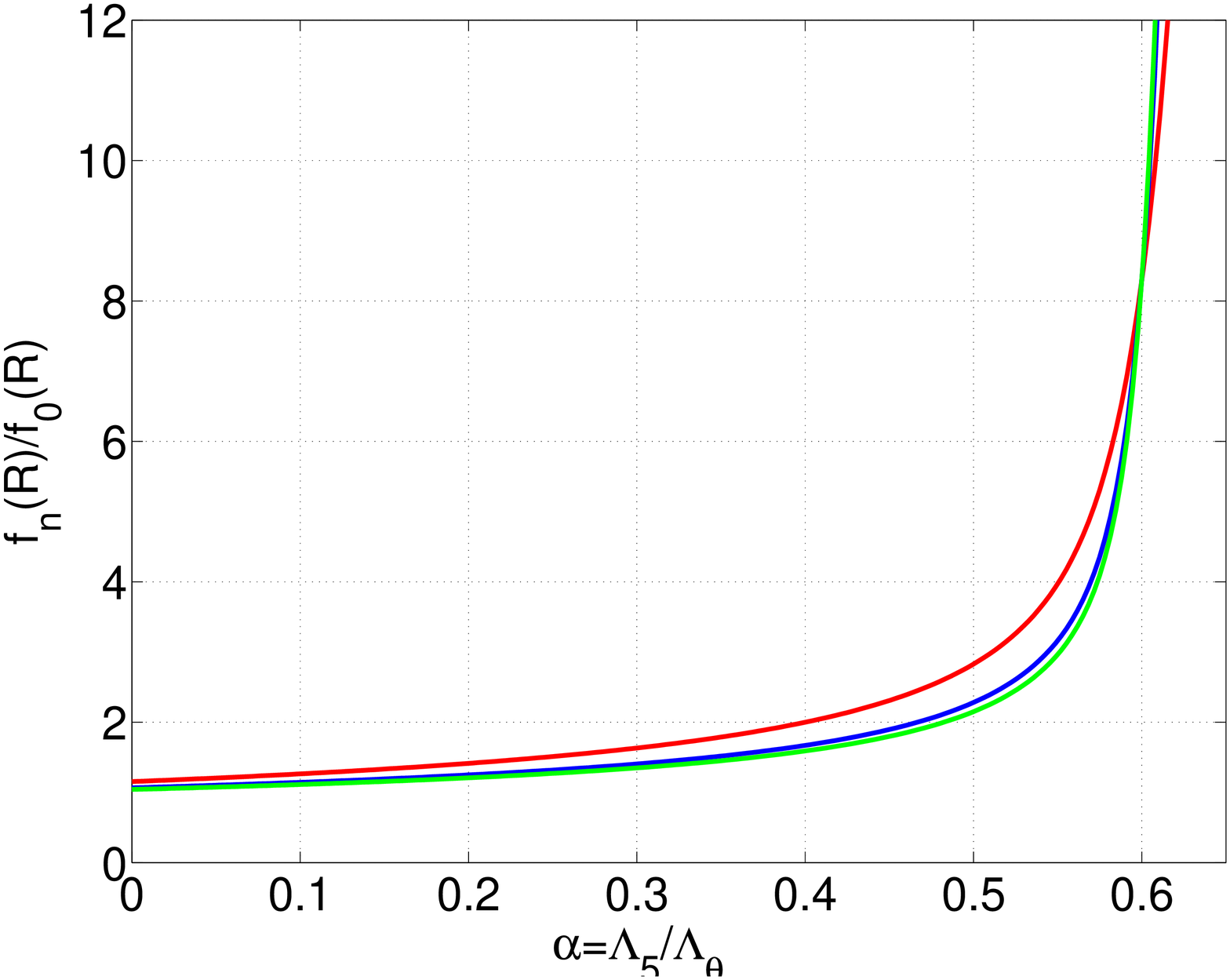}
\caption{The relative coupling of the $l_i=0$ KK gauge modes with a IR localised Higgs / fermion (on the left) and a UV localised fermion (on the right) in 6D (red), 8D (blue) and 10D (green). The lower graph is an enhancement of the couplings of IR localised fermions for small $\alpha$. Here $\Omega\equiv e^{kR}=10^{15}$ and $M_{KK}\equiv\frac{k}{\Omega}=1$TeV. }
\label{couplingfig }
\end{center}
\end{figure}

As was shown in \cite{Archer:2010hh} the scale of EW corrections is largely dependent on the couplings of the KK gauge modes with the Higgs and the fermion zero modes, relative to those of the gauge zero mode. Hence we define the relative gauge Higgs coupling to be
\begin{equation}\label{FN}
F_n\equiv \frac{f_n(R)\Theta_n(R_\theta)}{f_0(R)\Theta_0(R_\theta)}.
\end{equation}   
Note that although in order to resolve the hierarchy problem we require that the Higgs is localised close to $r=R$, there is still some freedom in choosing where in $\theta$ to locate the 3 brane. Alternatively if the Higgs is only localised w.r.t $r$, as explained in section \ref{EWconst}, the $l_i\neq 0$ modes will not contribute to EW observables at tree level. The relative gauge Higgs coupling for the $l_i=0$ modes have been plotted in figure \ref{couplingfig }.

Here we only consider fermions localised to either the UV 3 brane or the IR 3 brane. Hence the relative gauge fermion coupling is just given as
\begin{equation}\label{Fpsi}
F_\psi^{(n)}=F_n \qquad \mbox{for IR fermions,}\qquad F_\psi^{(n)}= \frac{f_n(0)\Theta_n(0)}{f_0(R)\Theta_0(0)}\qquad \mbox{for UV fermions}. 
\end{equation}
Since the tree level EW constraints are only concerned with the fermion zero mode, the relevant gauge fermion couplings (i.e. the gauge coupling with the $l_i=0$ fermion modes) would not change if the fermions were also only localised w.r.t $r$ because the fermion profile would be flat w.r.t $\theta$. Also note that the case in which the fermions are localised on the UV brane should be taken as an approximation of a model with bulk fermions localised towards the UV brane. Again these have been plotted for the $l_i=0$ modes in figure \ref{couplingfig }.  
   
\subsection{The Gauge Scalars}
We now move to consider the scalar components of the gauge fields. For more work on gauge scalars in 6D see \cite{Burdman:2005sr, Cacciapaglia:2009pa} or in D dimensions \cite{McDonald:2009hf}. Firstly since the mass terms of equations (\ref{A5EOM}) and (\ref{AiEOM}) are gauge dependant the physical gauge scalars are not just $A_{5}$ and $A_i$. To eliminate the gauge dependant part we take $\partial_i$(\ref{A5EOM})$-\partial_5$(\ref{AiEOM}) and $\partial_i$(\ref{AiEOM})$-\partial_j$(\ref{AiEOM}) to get 
\begin{eqnarray}
\partial_5^2\Phi_{i5}-(6k+\delta J-2J)\partial_5\Phi_{i5}+e^{2Jr}\sum_{j=1}^{\delta}\left (\partial_j^2\Phi_{i5}-(2k-2J)\partial_j\Phi_{ij}\right )\nonumber\\
+2k(4k+\delta J-2J)\Phi_{i5}+e^{2kr}m_n^{\Phi_{i5}\,2}\Phi_{i5}=0\label{EOMi5}\\
\partial_5^2\Phi_{ij}-(4k+\delta J-2J)\partial_5\Phi_{ij}+e^{2Jr}\sum_{k=1}^\delta \partial_k^2\Phi_{ij}+e^{2kr}m_n^{\Phi_{ij}\,2}\Phi_{ij}=0\label{EOMij}.
\end{eqnarray}
Where we have defined the physical gauge scalars to be
\begin{equation*}
\Phi_{i5}\equiv \partial_iA_5-\partial_5A_i\qquad\mbox{and}\qquad \Phi_{ij}\equiv \partial_iA_j-\partial_jA_i  
\end{equation*}
with masses
\begin{equation*}
\partial_\mu\partial^\mu\Phi_{i5}=-m_n^{\Phi_{i5}\,2}\Phi_{i5}\qquad\mbox{and}\qquad\partial_\mu\partial^\mu\Phi_{ij}=-m_n^{\Phi_{ij}\,2}\Phi_{ij}. 
\end{equation*}
At first sight one may be alarmed by having gone from $1+\delta$ gauge field components to $\left (\begin{array}{c}\delta+1\\2\end{array}\right )$ combinations of physical gauge scalars but of course they are not all independent and are related by
\begin{equation}\label{phimix}
\partial_5\Phi_{ij}=\partial_j\Phi_{i5}-\partial_i\Phi_{j5}.
\end{equation}  
Hence there is in fact $\delta$ physical gauge scalars formed from the mixing of $\Phi_{ij}$ and $\Phi_{i5}$. The remaining one component forms a KK tower of Goldstone bosons which are then `eaten' by the KK gauge fields,
\begin{equation*}
\varphi=e^{-(2k-2J)r}\left (-(2k+\delta J)e^{-2Jr}A_5+e^{-2Jr}\partial_5A_5+\sum_{i=1}^{\delta}\partial_iA_i\right )
\end{equation*}
with the equation of motion
\begin{equation*}
\partial_\mu \partial^\mu\varphi-\xi\left (e^{-2kr}\partial_5^2\varphi-(2k+\delta J)e^{-2kr}\partial_5\varphi+e^{-(2k-2J)r}\sum_{i=1}^{\delta}\partial_i^2\varphi=0\right ).
\end{equation*}
Note if we define the Goldstone bosons mass by $\partial_\mu \partial^\mu\varphi=-m^2_\varphi\varphi$ then in the unitary gauge $\xi\rightarrow\infty$, the Goldstone bosons become infinitely heavy and hence are unphysical.

We define the usual KK decomposition for $\Phi_{i5}$
\begin{displaymath}
\Phi_{i5}=\sum_{n}\Phi^{(n)}_{i5}(x^{\mu})f_n^{\Phi_{i5}}(r)\Theta_n^{\Phi_{i5}}(\theta_1,\dots,\theta_\delta)
\end{displaymath}
and likewise for $\Phi_{ij}$. Restricting ourselves to toroidal additional dimension with $\partial_j^2\Phi_{i5}=-\frac{l_j^2}{R_\theta^2}\Phi_{i5}$. Then the profiles of the KK modes with $l_i=0$ (i.e. flat w.r.t. $\theta_i$) can be obtained by solving (\ref{EOMij}), to get 
\begin{displaymath}
f_n^{\Phi_{ij}}(l_i=0)=Ne^{\frac{(4k+\delta J-2J)r}{2}}\left (\mathbf{J}_{\frac{(2-\delta)J-4k}{2k}}\left (\frac{m_n^{\Phi_{ij}}e^{kr}}{k}\right )+\beta\mathbf{Y}_{\frac{(2-\delta)J-4k}{2k}}\left (\frac{m_n^{\Phi_{ij}}e^{kr}}{k}\right )\right ).
\end{displaymath}
Likewise (\ref{EOMi5}) can be solved to give
\begin{displaymath}
f_n^{\Phi_{i5}}(l_i=0)=Ne^{\frac{(6k+\delta J-2J)r}{2}}\left (\mathbf{J}_{\frac{|(\delta-2)J+2k|}{2k}}\left (\frac{m_n^{\Phi_{i5}}e^{kr}}{k}\right )+\beta\mathbf{Y}_{\frac{|(\delta-2)J+2k|}{2k}}\left (\frac{m_n^{\Phi_{i5}}e^{kr}}{k}\right )\right ).
\end{displaymath}
The physical modes would then be a mixture of $\Phi_{ij}$ and $\Phi_{i5}$. Note that (\ref{phimix}) would almost certainly mix $l_i=0$ modes and $l_i\neq0$ modes and hence to compute the KK spectrum it is neccesary to carry out a full analysis of all modes. Such an analysis would be dependent  on the dimensionality of the additional dimensions and is beyond the scope of this work. Here we merely wish to emphasise that the phenomenology of such gauge scalars would probably be very different from that of normal scalars. For example, analogous to the spin connection of fermions in warped spaces, the warping of the space generates an effective mass for the gauge scalars and hence the zero mode of $\Phi_{i5}$ would not be constant w.r.t $r$,
\begin{displaymath}
f_0^{\Phi_{i5}}=N(e^{2kr}+\beta e^{(4k+\delta J-2J)r}).
\end{displaymath}  
Bearing in mind the physical zero modes would involve a mixing between $\Phi_{i5}^{(0)}$ and $\Phi^{(0)}_{ij}$ with $f_0^{\Phi_{ij}}=N$, one would not expect a flat zero mode. 

However in any phenomenologically realistic scenario one would not expect massless (or low mass) charged scalars and hence would expect $(-+)$ or $(--)$ BC's. Imposing such BC's would mean that such gauge scalars do not contribute at tree level to EW observables. So here we will leave a full analysis of the gauge scalars to future work. Although it is worth pointing out that with strict Dirichlet BC's and the Higgs and fermions localised to 3 branes, it would appear that the lowest gauge scalar KK mode of an abelian gauge field is stable. One could speculate that this could be a potential dark matter candidate and work has been done in this direction in \cite{Cacciapaglia:2009pa}. But such a particle could easily be destabilised by for example coupling it to any bulk field or relaxing the assumption of infinitely thin branes. The non-abelian gauge scalar would decay via, for example, the $g\varepsilon^{abc}A_\mu^bA_i^c(\partial^\mu A_i^a-\partial_iA^{\mu\;a})$ term in the Lagrangian.           
 
\section{Electroweak constraints}\label{EWconst}

\begin{figure}[!t]
\begin{center}
\begin{tabular}{c}
\includegraphics[height=2.8in]{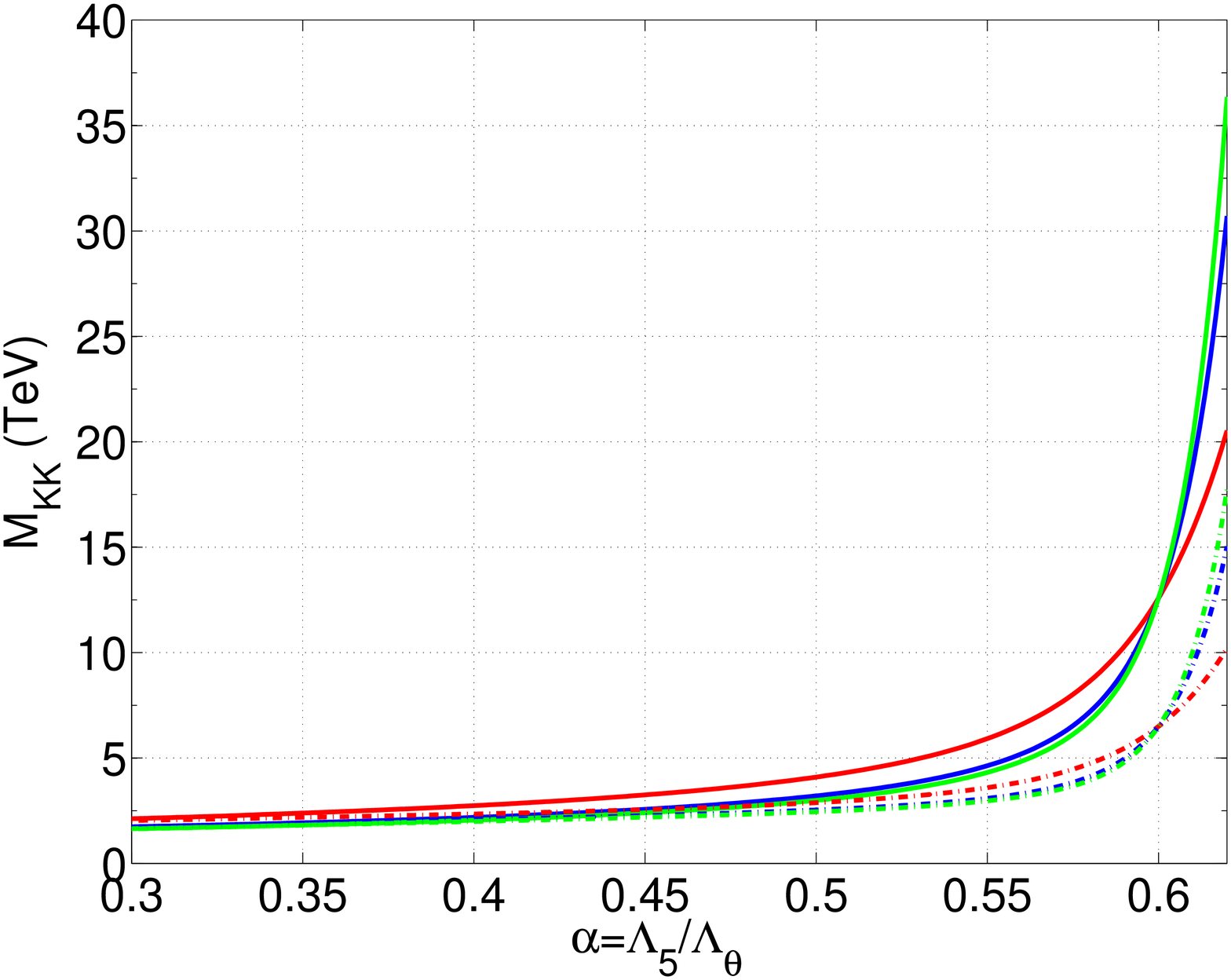}\\
\includegraphics[height=2.8in]{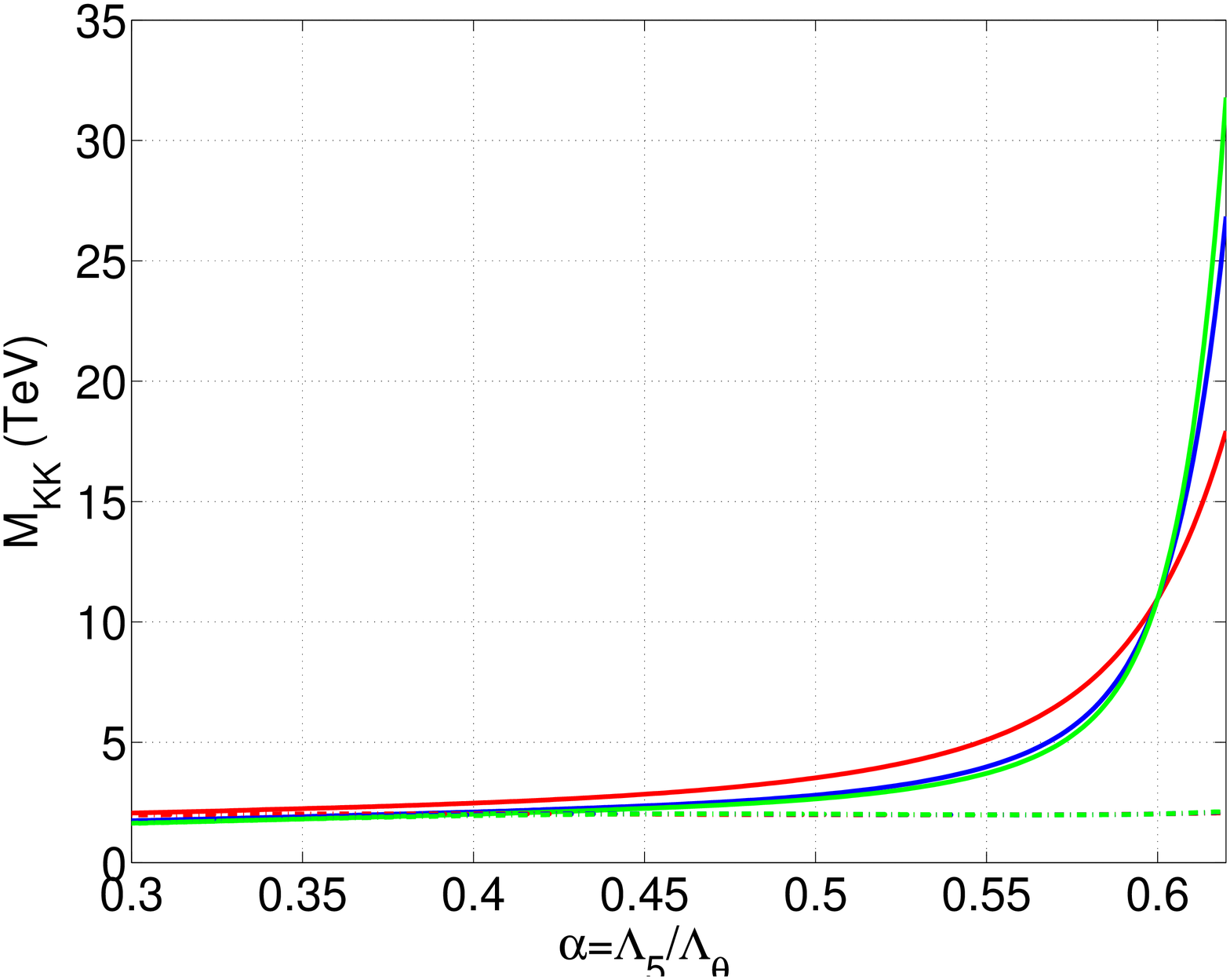}\\
\end{tabular}
\caption{The lower bound on $M_{KK}\equiv\frac{k}{\Omega}$ arising from the EW observable $s_Z^2$. With a bulk $SU(2)\times U(1)$ gauge symmetry (top) and a bulk $SU(2)_R\times SU(2)_L\times U(1)$ custodial symmetry (bottom). Here the fermions are localised to the IR brane (solid line) or the UV brane (dot-dash line) while the gauge fields propagate in 6D (red), 8D (blue) and 10D (green). The Higgs is localised w.r.t $r$ such that $\Omega=10^{15}$.  }
\label{SZfig}
\end{center}
\end{figure}

\begin{figure}[h]
\begin{center}
\begin{tabular}{cc}
\includegraphics[width=3.4in]{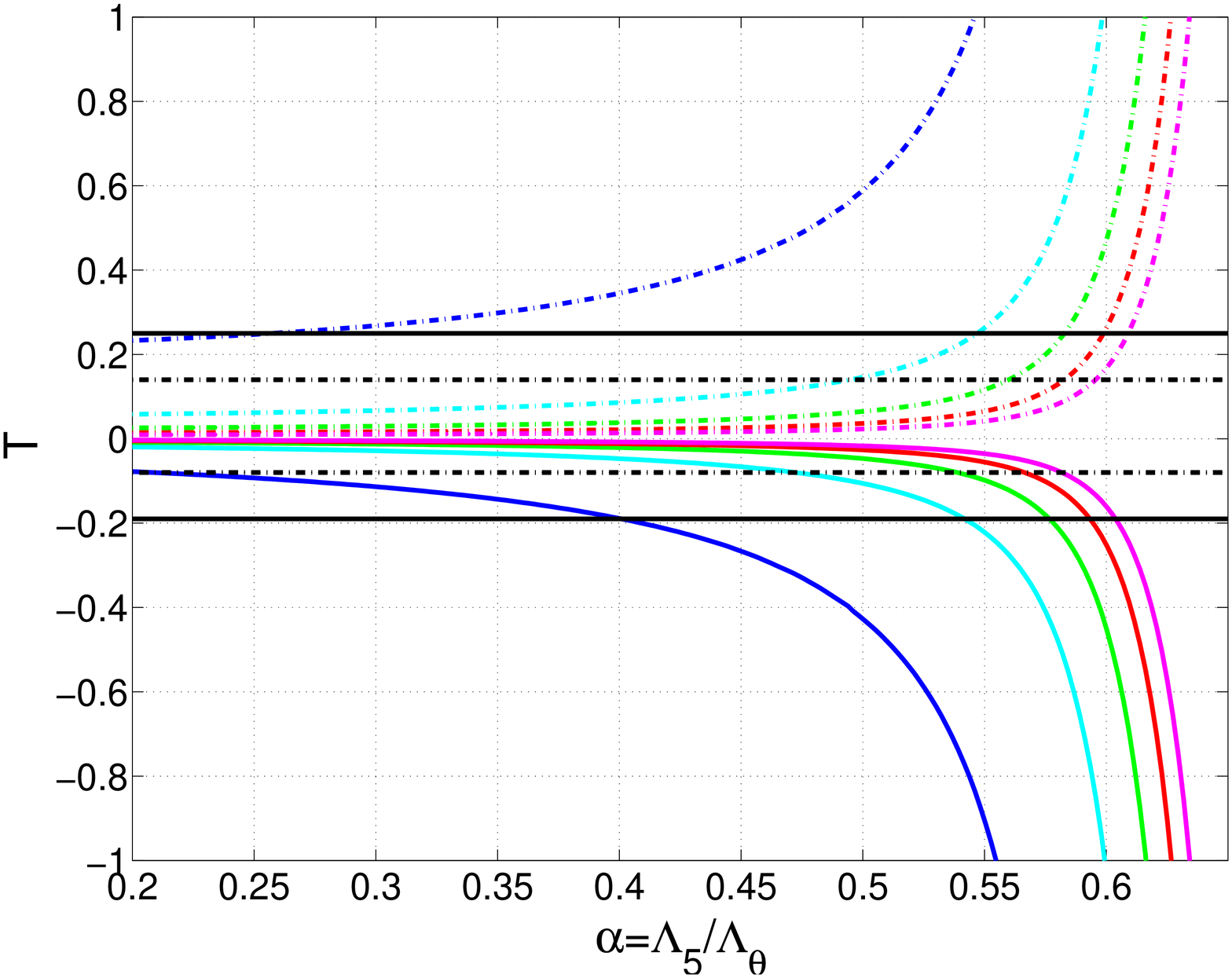}&
\includegraphics[width=3.4in]{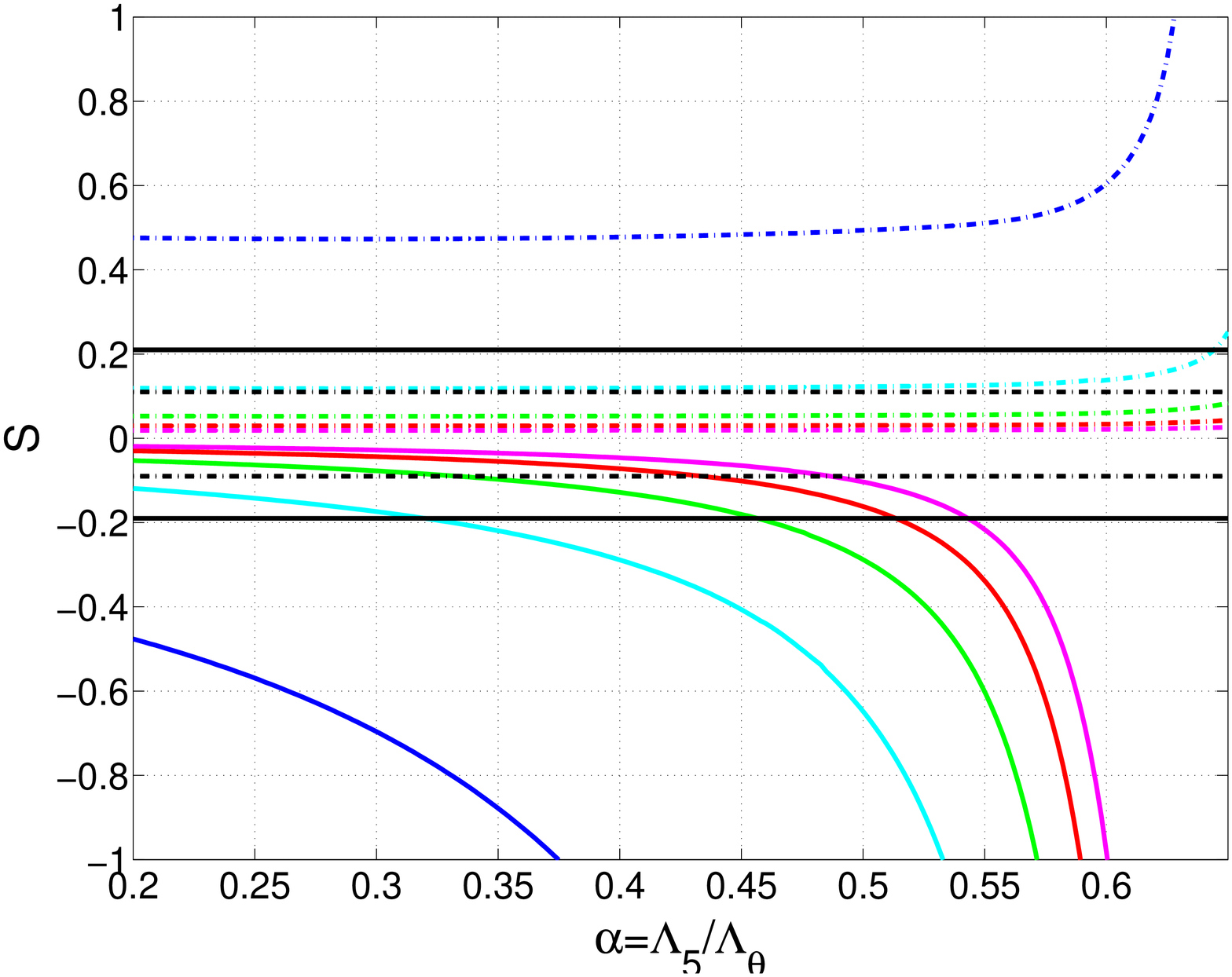}\\
\end{tabular}
\caption{The tree level contribution to the S and T parameters in six dimensional models, with out a custodial symmetry, where the KK scale ($M_{KK}\equiv\frac{k}{\Omega}$) is 1 TeV (blue), 2 TeV (cyan), 3 TeV (green), 4 TeV (red) and 5 TeV (magenta). The fermions are localised on 3 branes in the IR, $r=R$ (solid lines) or on 3 branes in the UV $r=0$. (dot dash lines). In black are the $1\sigma$ and $2\sigma$ bounds ($M_H=117$ GeV) on S and T \cite{Nakamura:2010zzi}  $\Omega\equiv e^{kR}=10^{15}$.  }
\label{fig:STnoCustodial}
\end{center}
\end{figure}

\begin{figure}[h]
\begin{center}
\begin{tabular}{cc}
\includegraphics[width=3.4in]{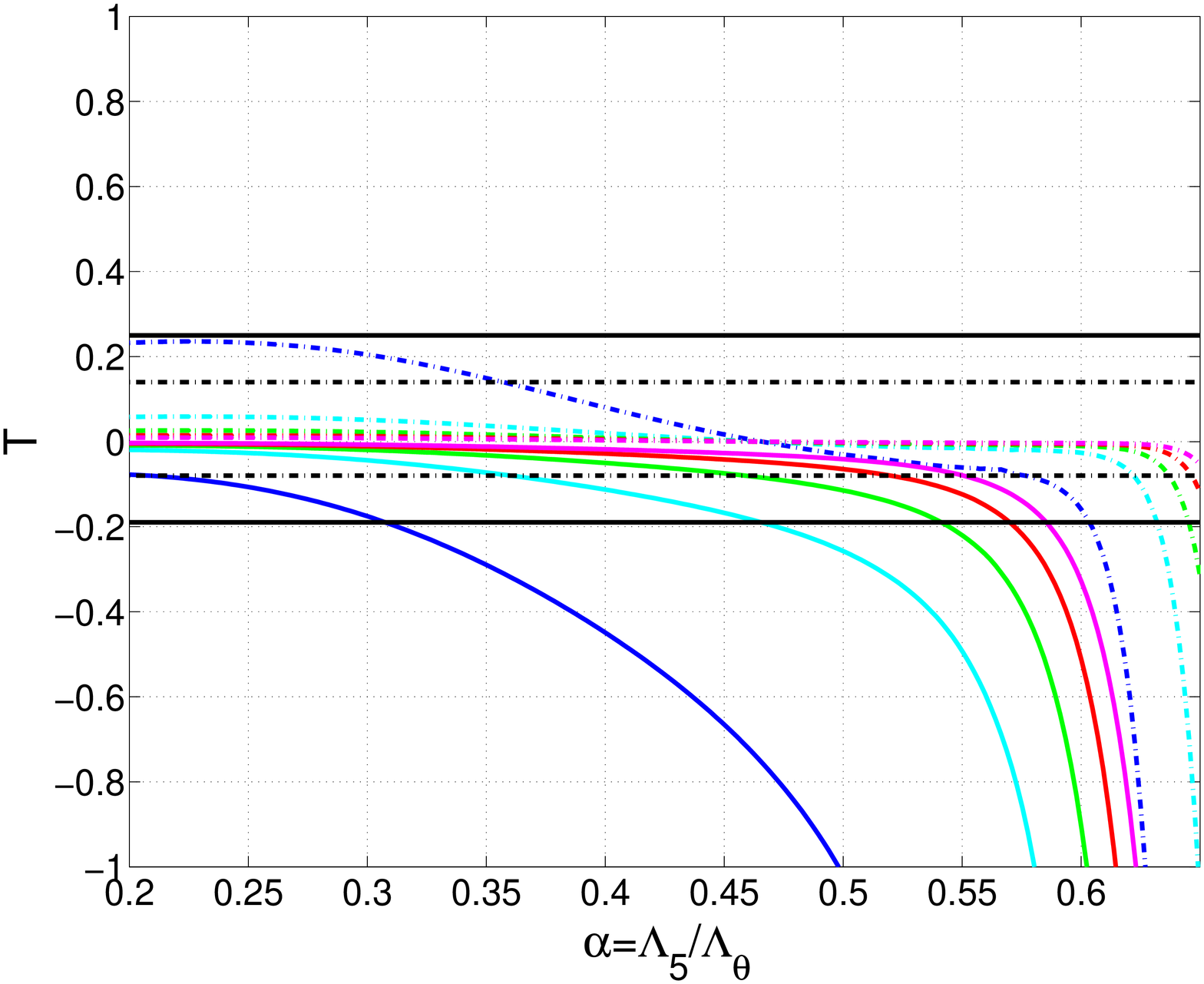}&
\includegraphics[width=3.4in]{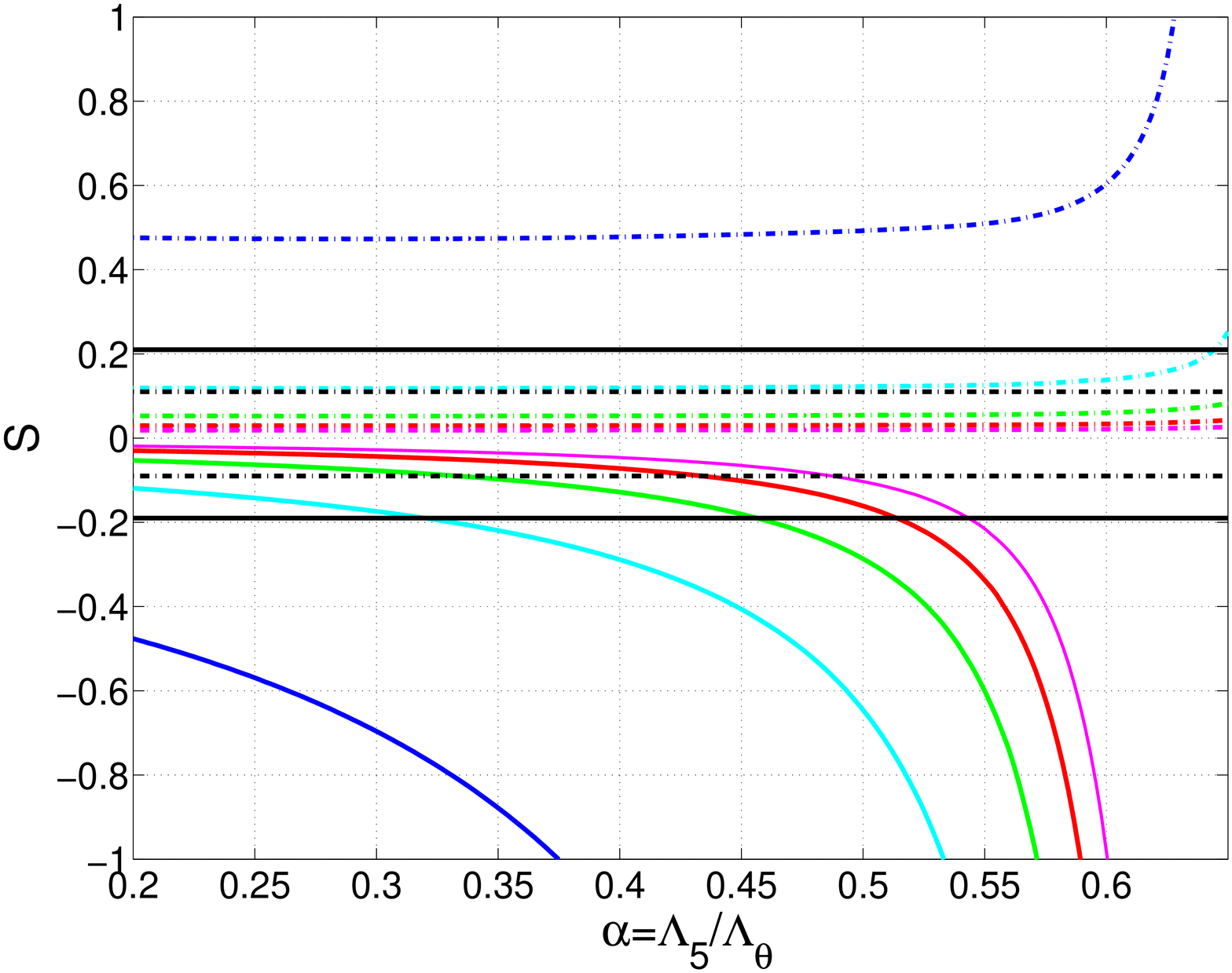}\\
\end{tabular}
\caption{As in figure \ref{fig:STnoCustodial} the tree level S and T parameters but now with a bulk $SU(2)_R\times SU(2)_L\times U(1)$ custodial symmetry.}
\label{fig:STparametersCustodial}
\end{center}
\end{figure}

Having carried out the KK decomposition of a bulk gauge field we now move to estimate the size of the lower bound on the KK mass scale arising from EW constraints. Ideally one would carry out a full $\chi^2$ test for each possible geometry. However this would be computationally cumbersome and hence here we take two different but simpler approaches. Firstly in \cite{Archer:2010hh} the input parameters were fixed by comparison with the three most precisely measured EW observables ($\alpha$, $G_f$ and $\hat{M}_Z$). Then eight LEP 1 observables were computed at tree level and a constraint was arrived at by ensuring that the deviation between tree level SM value and the computed value was within $2\sigma$ of the experimental result. This is essentially equivalent to assuming that any deviation between the SM and experiment will be dominated by beyond the SM tree level effects. It was found in all cases that the tightest constraint came from the weak mixing angle, $s_Z^2=0.23119\pm0.00014$ \cite{Amsler:2008zzb}, which is defined to be the ratio of the couplings at the Z pole. Hence here we compute the constraint from just this observable but for two models. One with a  bulk $SU(2)\times U(1)$ EW sector and the other with a bulk $SU(2)_R\times SU(2)_L\times U(1)$ custodial symmetry. We also consider the two `extremal' cases of fermions localised on the IR brane and fermions localised on the UV brane. One would anticipate that if the fermions were free to propagate in the bulk then, at tree level, the constraints would lie between these two values. The weak mixing angles are computed in (\ref{szNoCust}) and (\ref{SZapp}) and the constraints are plotted in figure \ref{SZfig}.    
 
The second more conventional approach is to parameterise the EW constraints in terms of the S and T parameters \cite{Peskin:1991sw}.  Note that such a parameterisation assumes that one can absorb the non oblique corrections which requires that one assumes universal gauge fermion coupling. These parameters are calculated in (\ref{eq:ST}) and (\ref{eqn:STcust}) and plotted in figure \ref{fig:STnoCustodial} and figure \ref{fig:STparametersCustodial}. These have been compared to bounds taken from \cite{Nakamura:2010zzi} of $\rm{S}=0.01\pm0.1$ and $\rm{T}=0.03\pm0.11$ ($M_H=117$ GeV) although a recent fit has put the bounds at $\rm{S}=0.02\pm0.11$ and $\rm{T}=0.05\pm0.12$ ($M_H=120$ GeV) \cite{Haller:2010zb}. We have just plotted the case for six dimensions where, for $\alpha<0.6$, one would expect the largest constraints as confirmed by figure \ref{SZfig}.  

In both approaches we have not plotted the constraints for $\alpha>0.6$ since here the gauge fields would become strongly coupled and this perturbative analysis would not be valid. However one would anticipate the constraints being large. Firstly it is reassuring to note that the constraints from the two approaches are in reasonably good, but not exact, agreement. When $\alpha=0.6$, the internal space is unwarped and  hence not scaling the couplings. So one would expect to obtain similar constraints to those of the 5D RS model. It is found, for KK scales greater than about 1 TeV, that the constraints from the S and T parameters are in good agreement with those found in \cite{Csaki:2002gy, Carena:2007ua, Casagrande:2008hr}. Bearing in mind the differences in the approaches and the sensitivity to the KK scale, the level of agreement is quite surprising. The results are of course in agreement with figure \ref{couplingfig } in the fact that, as $J$ becomes increasingly negative, the gauge Higgs coupling is volume suppressed and the EW constraints are suppressed. 

Readers may be surprised by the potentially large contribution to the T parameter in models with a custodial symmetry. As explained in the appendix, in models with an unbroken custodial symmetry, the coefficient of the operator $|HDH|^2$ would be zero. However one is forced, in breaking the custodial symmetry, by imposing different boundary conditions on the $SU(2)_R$ and $SU(2)_L$ fields. This leads to a difference in the couplings and masses, see figure \ref{gaugemassesfig}, which in turn leads to a positive T parameter for small $\alpha$. There is also a negative contribution from the absorption of the non oblique corrections, as well as a differing correction to the KK modes masses (i.e. the $m_w^2(F_n^2-1)$ term), which are both only partially protected by the custodial symmetry. This leads to a negative contribution to the T parameter which is enhanced for large $\alpha$. Having said that the dominant constraint still arises from the S parameter, for both IR and UV localised fermions and hence is dependent on the $F_nF_\psi^{(n)}$. This can also be seen in (\ref{SZapp}).

When there is no custodial symmetry and the fermions are sitting towards the UV brane the constraints arise from the T parameter or, in (\ref{szNoCust}), the $F_n^2$ terms. The point is that in this model there appears to be an overall lower bound, of around $M_{KK}\gtrsim 2$ TeV, corresponding to
\begin{displaymath}
\frac{F_n^2}{m_n^2},\;\frac{F_nF_\psi}{m_n^2}\sim\frac{1}{m_n^2}.
\end{displaymath}  
To reduce the constraints any lower one would need the KK modes of the gauge field to be more weakly coupled, to the Higgs and the fermions, than the zero mode. Although it should be noted that the perturbative approach taken here would probably break down at this low a KK scale. None the less here it does not seem to be possible to decouple the KK modes using the warping of the internal space. However one can use this warping to suppress the couplings and reduce the constraints, regardless of the bulk gauge symmetry or the location of the fermions.

\section{Conclusion}
In this paper we have considered what is arguably the simplest $D$ dimensional extension of the RS model. That in which the space is warped w.r.t a single direction $r$ and the warping has resulted from an anisotropic bulk cosmological constant. It has been very much a `bottom up' classical approach in which we have neglected back reactions from bulk fields. Nonetheless solving the Einstein equations yield essentially three interesting classes of solutions. Those in which the warp factor of the additional dimensions is increasing towards the IR, those in which it is decreasing and the particular case in which the additional dimensions are not warped. It is well known that allowing gauge fields to propagate in these extra dimensions will give rise to corrections to the EW observables at tree level arising from the $D-4$ KK towers of gauge bosons. Carrying out the KK decomposition we find that if we take the radius of these $D>5$ extra dimensions to be $\sim M_P^{-1}$ then in spaces in which the additional dimensions are decreasingly warped towards the IR ($J> 0$) then $D-5$ of the KK towers will not begin until a scale well above that of $M_{KK}$. On the other hand if the warping is increasing towards the IR ($J<0$) then the spacing of the $D-5$ KK towers would be small and hence one would see a fine spacing in the KK mass spectrum at the KK scale. Likewise if the warping of the extra dimensions is very small ($J\sim 0$) one would anticipate a conventional spacing in the mass spectrum. Further still we find after separating the unphysical gauge dependent part of the gauge field from the remaining physical scalar part of the higher dimensional vector field, that the resulting gauge scalar has a different wavefunction from that of a normal higher dimensional scalar field. Notably that the zero mode would no longer be necessarily flat. Requiring that there are no massless gauge scalars would require Dirichlet boundary conditions and could potentially lead to the lowest KK mode of a abelian gauge scalar being stable. Although such boundary conditions would mean the gauge scalars would decouple from the electroweak sector at tree level.

It has been previously shown that the size of the EW corrections is determined largely by the size of the KK gauge-Higgs coupling relative that of the zero mode. It has also been shown that in spaces which are warped increasingly (decreasingly) towards the IR brane the volume of the $D>5$ extra dimensions scale the KK modes differently to that of the zero mode, resulting in the relative coupling being suppressed (enhanced). This has been confirmed for the spaces studied in this paper (see figure \ref{couplingfig }) and the resulting EW constraints are reduced (see figure \ref{SZfig}, \ref{fig:STnoCustodial} and \ref{fig:STparametersCustodial}). However this assumes the Higgs is localised only w.r.t $r$ and hence the $D-5$ additional KK towers are orthogonal to the SM gauge fields (assumed to be the zero modes). On the other hand if the Higgs was localised to a 3 brane these extra KK modes would contribute to EW observables and the EW constraints would not be perturbatively computable. However, with a Higgs localised on a codimension one brane, we find a lower bound on the KK mass scale of $M_{KK}\gtrsim 2$ TeV regardless of whether there is a bulk custodial symmetry or not. 

The model outlined in this paper can only be considered as a toy model for a number of reasons. Firstly it is very much a `bottom up model' in so much as it makes no contact with string theory and still requires a UV cut off. Secondly the classical solutions to the Einstein equations considered here are by no means the most general solutions and really one should work with an ansatz warped in multiple directions. We have also neglected the back reaction of the brane tension and bulk fields on the solution and the model potentially suffers from singular branes. Thirdly due to the difficulties in producing a 4D chiral theory, here the fermions are strictly localised to 3 branes. As mentioned earlier many of these problems have known solutions and one can speculate about how these results would change if we implemented these solutions. For example if one worked with just codimension one branes, the fermions would be forced into the higher dimensional space and their couplings would result from overlap integrals including the warping of the internal space. Hence, as in the 5D case, one can view the couplings of fermions localised on 3 branes as the extremal case of bulk fermions. So one would anticipate the constraints for bulk fermions lying between these two cases considered, i.e. between the IR and UV localised fermions. Likewise  it is hard to see how, considering solutions that avoid the singular branes, would forbid the class of solutions corresponding to $\alpha<0.6$.  

The central point this paper wishes to highlight  is that there are a number of interesting phenomenological features that arise from the warping of the internal manifold of spaces of more than five dimensions. In particular the additional splitting in the KK mass spectrum, the enhancement or suppression of the KK modes coupling's and the difference between gauge scalars and normal scalars. Here we have demonstrated that these differences dramatically alter the constraints arising from EW precision tests but one would anticipate that these effects would alter much of the phenomenology of the extra dimensional models. Clearly it is necessary to properly understand these deviations from the 5D theory before we can offer any interpretations of signals at the LHC.

\section*{Acknowledgements}
P.R.A.~was supported by the Science and Technology Facilities Council.\newline
S.J.H.~was supported by the Science and Technology Facilities Council [grant number ST/G000573/1].

\appendix

\section{Tree Level Electroweak Constraints from models without a Custodial Symmetry}
Here we will briefly derive the tree level constraints on the EW observable, but we refer the reader to \cite{Csaki:2002gy, Carena:2002dz, Archer:2010hh, Casagrande:2008hr,Goertz:2008vr} for a more detailed account. If we start by considering the EW sector post KK decomposition but pre spontaneous symmetry breaking,
\begin{eqnarray}
S=\sum_n\int d^4x\Big [ -\frac{1}{4}A^{(n)a}_{\mu\nu}A^{(n)a\mu\nu}-\frac{1}{4}B_{\mu\nu}^{(n)}B^{\mu\nu(n)}+\frac{1}{2}m_{n}^{2}A_\mu^{(n)}A^{\mu(n)}+\frac{1}{2}m_n^2B^{(n)}_\mu B^{(n)\mu}\nonumber\\
+\sum_{\Psi}\;i\bar{\Psi}\gamma^\mu D_\mu\Psi+|D_\mu\Phi|^2 +V(\Phi)\Big ],
\end{eqnarray}
where we have denoted the $SU(2)_L$ and $U(1)_X$ fields as $A_\mu^a$ and $B_\mu^a$. The covariant derivative will then be
\begin{equation}
D_\mu=\partial_\mu+\sum_n\left (-igf_{n}\Theta_nA_\mu^{a(n)}\tau^a-ig^{\prime}f_{n}\Theta_nB_\mu^{(n)}\right ).\label{CovDerivative}
\end{equation}
where, through a slight abuse of notation, $f_n$ and $\Theta_n$ denotes an overlap integral between gauge fields and the fields they are coupling to. For example if the fermions were localised to a 3 brane at $r=R$ and $\theta=\theta_{\rm{ir}}$ then $f_n\Theta_n=f_n(R)\Theta_n(\theta_{\rm{ir}})\equiv f_\psi^{(n)}$. As explained in \cite{Archer:2010hh}, the gaining of a vev by the Higgs will cause the gauge KK modes to mix together. If, on the other hand, the Higgs is localised to a codimension one brane then the $l_i\neq 0$ gauge modes will not mix with the gauge zero mode, due to (\ref{gaugeOrthog}) and hence not contribute, at tree level, to the EW constraints. So if the Higgs gains a vev, $\Phi\rightarrow\frac{1}{\sqrt{2}}\left (\begin{array}{ c}0\\v+H\end{array}\right )$ and one makes the usual field redefinitions 
\begin{displaymath}
W_\mu^{\pm(n)}=\frac{1}{\sqrt{2}}\left (A_\mu^{1\;(n)}\mp iA_\mu^{2\;(n)}\right ),\quad Z_\mu^{(n)}=cA_\mu^{3\;(n)}-sB_\mu^{(n)},\quad A_\mu^{(n)}=sA_\mu^{3\;(n)}+cB_\mu^{(n)}
\end{displaymath}
where $s=\frac{g^{\prime}}{\sqrt{g^{\prime\;2}+g^2}}$ and $c=\frac{g}{\sqrt{g^{\prime\;2}+g^2}}$. Then the gauge mass term will now be
\begin{eqnarray}
\mathcal{L}\supset\sum_{n,m}\left (m_n^2\delta_{mn}+\frac{g_5^2v^2}{4}f_nf_m\Theta_n\Theta_m\right )W_\mu^{+(n)}W^{-(m)\mu}\hspace{6cm}\nonumber\\
+\left (\frac{m_n^2}{2}\delta_{mn}+\frac{(g_5^2+g_5^{\prime 2})v^2}{8}f_nf_m\Theta_n\Theta_m\right )Z_\mu^{(n)}Z^{\mu (m)}\label{gaugeeigenstate}
\end{eqnarray}
Here it is convenient to define the quantities $m_z^2\equiv\frac{(g_5^2+g_5^{\prime 2})v^2}{4}f_0^2\Theta_0^2$ and $m_w^2\equiv\frac{g_5^2v^2}{4}f_0^2\Theta_0^2$ as well as referring to the relative couplings defined in (\ref{FN}) and (\ref{Fpsi}). One can then rotate (\ref{gaugeeigenstate}) into the mass eigenstate by acting with a unitary matrix approximately given by
\begin{equation}
\left (U_{W/Z}\right )_{nn}\approx 1\quad\mbox{and}\quad \left (U_{W/Z}\right )_{mn}\approx-\frac{m_{w/z}^2F_nF_m}{m_m^2-m_n^2+m_{w/z}^2(F_m^2-F_n^2)}+\mathcal{O}(m_n^{-4}) \label{UnitaryMatrix}
\end{equation} 
such that the mass eigenvalues are given by
\begin{equation}
\left (M_{W/Z}^2\right )_{nn}\approx m_n^2+m_{w/z}^2F_n^2-\sum_{m\neq n}\frac{\left (m_{w/z}^2F_nF_m\right )^2}{m_m^2-m_n^2+m_{w/z}^2\left (F_m^2-F_n^2\right )}+\mathcal{O}(m_n^{-4}).\label{massgaugestate}
\end{equation}
The resulting $W/Z$ zero mode profile, in the mass eigenstate, will no longer be flat. It is important to note that if the Higgs was not localised on a codimension one brane, then it would be necessary to include all the $l_i\neq 0$ modes in these summations. Summations over more than one KK tower are typically divergent, when $m_n^2\sim n^2$ and hence it is necessary to include a cut off in the KK number \cite{Appelquist:2000nn}. In other words if the Higgs is not localised to a codimension one brane then the EW corrections become dependent on an arbitrary cut off and it is not possible to compute a meaningful bound, using a perturbative approach. 

In \cite{Archer:2010hh}, as an alternative to a full $\chi^2$ test, the input parameters were fixed by comparison with the three observables with the least experimental error, 
\begin{eqnarray*}
\hat{M}_Z^2=\left (M_Z^2\right )_{00}\approx m_z^2\left (1-\sum_{n=1}\frac{m_z^2F_n^2}{m_n^2-m_z^2(F_n^2-1)}\right ),\label{MZobsC}\\
\sqrt{4\pi\alpha(M_Z)}=gsf_\psi^{(0)},\label{alphaC}\\
4\sqrt{2}G_f=g^2(f_\psi^{(m)})^\dag\,(\mathcal{M}_{W}^2)_{mn}^{-1}f_\psi^{(n)},\label{GfC}
\end{eqnarray*}
where $\mathcal{M}_{W}^2$ is the mass matrix (\ref{gaugeeigenstate}) in the gauge eigenstate. With the input parameters range of observables were computed and it was typically found that the tightest constraint came from the weak mixing angle $s_{Z}^2=\frac{g^{\prime\;2}}{g^2+g^{\prime\;2}}$ measured at the Z pole, $s_Z^2=0.23119\pm 0.00014$\cite{Amsler:2008zzb}, which at tree level is given by
\begin{displaymath}
s_Z^2=\frac{\pi\alpha}{\sqrt{2}G_f}\frac{f_\psi^{(m)}(\mathcal{M}_W^2)^{-1}_{nm}f_\psi^{(n)}}{(f_\psi^{(0)})^2}=\frac{\pi\alpha}{\sqrt{2}G_f}\left (\frac{1}{m_w^2}+\sum_{n=1}\frac{\left(F_n-F_\psi^{(n)}\right )^2}{m_n^2}\right ).
\end{displaymath}
Then using the approximation that $m_w^2\approx m_z^2 (1-s_Z^2)$ one finds that
\begin{equation}
\label{szNoCust}
s_Z^2\approx s_p^2\left (1-\frac{c_p^2}{c_p^2-s_p^2}\sum_{n=1}\left [\frac{m_z^2F_n^2}{m_n^2}-\frac{m_w^2\left(F_n-F_\psi^{(n)}\right )^2}{m_n^2}\right ]+\mathcal{O}(m_n^{-4})\right ).
\end{equation} 
where $s_p^2=\frac{1}{2}\left (1-\sqrt{1-\frac{4\pi\alpha}{\sqrt{2}G_f\hat{M}_Z^2}}\right )$ and $c_p^2=1-s_p^2$.

Alternatively one can compute the $S$ and $T$ parameters \cite{Peskin:1991sw} defined such that, if the 4D effective Lagrangian is given by
\begin{eqnarray*}
\mathcal{L}=-\frac{1}{2}(1-\Pi_{WW}^{\prime})W_{\mu\nu}^+W^{\mu\nu}_--\frac{1}{4}(1-\Pi_{ZZ}^{\prime})Z_{\mu\nu}Z^{\mu\nu}-\frac{1}{4}(1-\Pi_{\gamma\gamma}^{\prime})A_{\mu\nu}A^{\mu\nu}-\Pi_{\gamma Z}^{\prime}A^{\mu\nu}Z_{\mu\nu}\\
+\left (\frac{g_4^2v^2}{4}+\Pi_{WW}(0)\right )W_\mu^+W^\mu_-+\frac{1}{2}\left (\frac{(g_4^2+g_4^{\prime\;2})v^2}{4}+\Pi_{ZZ}(0)\right )Z_\mu Z^\mu+\sum_\Psi i\bar{\Psi}\gamma^\mu D_\mu\Psi,
\end{eqnarray*}
then the $S$ and $T$ Parameters are defined by
\begin{eqnarray*}
S=16\pi \left (\Pi_{33}^{\prime}-\Pi_{3Q}^{\prime}\right ) \\
T=\frac{4\pi}{s^2c^2M_Z^2}\left (\Pi_{11}(0)-\Pi_{33}(0)\right )
\end{eqnarray*} 
Note that such a parameterisation assumes that there is no correction to the gauge fermion coupling (or non oblique corrections). If there are then one must either include additional operators \cite{Cacciapaglia:2006pk, Grojean:2006nn} or use a different parameterisation. In the above scenario one would receive non oblique correction once you have rotated into the mass eigenstate and acted (\ref{UnitaryMatrix}) on the covariant derivative (\ref{CovDerivative}). However if one assumes universal corrections to the gauge fermion couplings, or equivalently that all the fermions are sitting at the same point, then one can make a field redefinition to absorb the non oblique corrections \cite{Agashe:2003zs}
\begin{eqnarray*}
W_\mu^{(0)}\rightarrow \left (1+\sum_n\frac{m_w^2F_nF_\Psi^{(n)}}{m_n^2+m_w^2(F_n^2-1)}+\mathcal{O}(m_n^{-4})\right )W_\mu^{(0)}\\
Z_\mu^{(0)}\rightarrow \left (1+\sum_n\frac{m_z^2F_nF_\Psi^{(n)}}{m_n^2+m_z^2(F_n^2-1)}+\mathcal{O}(m_n^{-4})\right )Z_\mu^{(0)}. 
\end{eqnarray*}
This then implies that
\begin{eqnarray*}
\Pi_{11}(0)=\frac{\Pi_{WW}(0)}{g_4^2}\approx\frac{v^2}{4}\sum_n\frac{m_w^2(2F_nF_\psi^{(n)}-F_n^2)}{m_n^2+m_w^2(F_n^2-1)}\\
\Pi_{33}(0)=\frac{\Pi_{ZZ}(0)}{g_4^2+g_4^{\prime\;2}}\approx\frac{v^2}{4}\sum_n\frac{m_z^2(2F_nF_\psi^{(n)}-F_n^2)}{m_n^2+m_z^2(F_n^2-1)}\\
\Pi_{3Q}^{\prime}=\Pi_{\gamma Z}^{\prime}=0\\
\Pi_{11}^\prime=\frac{\Pi_{WW}^{\prime}}{g_4^2}\approx-\frac{v^2}{4}\sum_n\frac{2F_nF_\psi^{(n)}}{m_n^2+m_w^2(F_n^2-1)}\\
\Pi_{33}^\prime=\frac{\Pi_{ZZ}^{\prime}}{g_4^2+g_4^{\prime\;2}}\approx-\frac{v^2}{4}\sum_n\frac{2F_nF_\psi^{(n)}}{m_n^2+m_z^2(F_n^2-1)}
\end{eqnarray*}
hence the tree level contribution to the S and T parameters are
\begin{eqnarray}
S\approx -\frac{4M_Z^2c^2s^2}{\alpha}\sum_n\frac{2F_nF_\psi^{(n)}}{m_n^2+m_z^2(F_n^2-1)}+\mathcal{O}(m_n^{-4})\nonumber\\
T\approx\frac{1}{\alpha}\sum_n\left ( \frac{m_w^2(2F_nF_\psi^{(n)}-F_n^2)}{m_n^2+m_w^2(F_n^2-1)}-\frac{m_z^2(2F_nF_\psi^{(n)}-F_n^2)}{m_n^2+m_z^2(F_n^2-1)}+\mathcal{O}(m_n^{-4})\right ).\label{eq:ST}
\end{eqnarray}

\section{Tree level Electroweak Constraints from Models with a Custodial Symmetry.} \label{CustApp}
We will now move on to consider models with a bulk gauge symmetry $SU(2)_R\times SU(2)_L\times U(1)_X\times P_{LR}$. As shown in \cite{Agashe:2003zs} the additional $SU(2)_R$ gauge symmetry protects the T parameter and hence reduces the EW constraints on the theory. There has since been considerable work on these models, see for example \cite{Csaki:2003zu, Cacciapaglia:2006gp, Contino:2006qr, Carena:2006bn, Casagrande:2010si} although here we largely follow \cite{Albrecht:2009xr} again using the perturbative approach used in \cite{Archer:2010hh, Goertz:2008vr}.

Denoting the $SU(2)_R$, $SU(2)_L$ and $U(1)_X$ fields as $\tilde{A}_M^a$, $A_M^a$ and $X_M$. Here the discrete symmetry $P_{LR}$ ensures the $SU(2)$ couplings are the same, $g_L=g_R\equiv g$ while the $U(1)$ symmetry has a coupling $g^\prime$. It is assumed that some Higgs Mechanism on the UV brane breaks the $SU(2)_R\times U(1)_X\rightarrow U(1)_Y$ leading to the field redefinitions
\begin{displaymath}
\tilde{Z}_M=c^{\prime}\tilde{A}_M^3-s^{\prime}X_M\quad\mbox{and}\quad B_M=s^{\prime}\tilde{A}_M^3+c^{\prime} X_M. 
\end{displaymath}
Where $s^{\prime}=\frac{g^{\prime}}{\sqrt{g^2+g^{\prime\,2}}}$ and $c^{\prime}=\frac{g}{\sqrt{g^2+g^{\prime\,2}}}$. This leads to the BC's
\begin{displaymath}
\tilde{Z}_M:\;(-,+)\quad \tilde{A}^{1,2}_M:\;(-,+)\quad B_M:\;(+,+)\quad A^a_M:\;(+,+).
\end{displaymath}
The remaining symmetry is then broken by a Higgs (now a bi-doublet under $SU(2)_L\times SU(2)_R$) on the IR brane which gains a VEV at the EW scale, $\Phi\rightarrow\left(\begin{array}{cc}0 & -\frac{v+H}{2} \\\frac{v+H}{2} & 0\end{array}\right)$. This then leads to the usual EW symmetry breaking and the field redefinitions
\begin{eqnarray}
Z_\mu^{(n)}=cA_\mu^{3(n)}-sB_\mu^{(n)},\hspace{2.5cm} A_\mu^{(n)}=sA_\mu^{3(n)}+cB_\mu^{(n)},\nonumber\\
W_\mu^{\pm\,(n)}=\frac{1}{\sqrt{2}}\left (A_\mu^{1(n)}\mp iA_\mu^{2(n)}\right ),\qquad \tilde{W}_\mu^{\pm\,(n)}=\frac{1}{\sqrt{2}}\left (\tilde{A}_\mu^{1(n)}\mp i\tilde{A}_\mu^{2(n)}\right ).\nonumber
\end{eqnarray}     
Where $s=\frac{s^{\prime}}{\sqrt{1+s^{\prime\,2}}}$ and $c=\frac{1}{\sqrt{1+s^{\prime\,2}}}$. If we now carry out the KK decomposition, then the analogous mass term to (\ref{gaugeeigenstate}) of the gauge fields will be given by
\begin{equation*}
\left (\begin{array}{cccc}W_\mu^{+\,(0)}& W_\mu^{+\,(1)}&\tilde {W}_\mu^{+\,(1)}&\dots\end{array} \right )\mathcal{M}_{\rm{charged}}^2\left (\begin{array}{c}W_\mu^{-\,(0)}\\W_\mu^{-\,(1)}\\\tilde{W}_\mu^{-\,(1)}\\\vdots\end{array}\right )
\end{equation*}
\begin{equation*}
\frac{1}{2}\left (\begin{array}{cccc}Z_\mu^{(0)}& Z_\mu^{(1)}&\tilde {Z}_\mu^{(1)}&\dots\end{array} \right )\mathcal{M}_{\rm{neutral}}^2\left (\begin{array}{c}Z_\mu^{(0)}\\Z_\mu^{(1)}\\\tilde{Z}_\mu^{(1)}\\\vdots\end{array}\right )
\end{equation*}
where
\begin{equation}\label{Mcharge}
\mathcal{M}_{\rm{charged}}^2=\left(\scriptsize\begin{array}{cccccc}\frac{g^2v^2}{4}f_0f_0 & \frac{g^2v^2}{4}f_0f_1 & -\frac{g^2v^2}{4}f_0\tilde{f}_1 & \frac{g^2v^2}{4}f_0f_2 & -\frac{g^2v^2}{4}f_0\tilde{f}_2 & \cdots \\\frac{g^2v^2}{4}f_0f_1 & m_1^2+\frac{g^2v^2}{4}f_1f_1 & -\frac{g^2v^2}{4}f_1\tilde{f}_1 & \frac{g^2v^2}{4}f_1f_2 & -\frac{g^2v^2}{4}f_1\tilde{f}_2 & \cdots \\-\frac{g^2v^2}{4}f_0\tilde{f}_1 & -\frac{g^2v^2}{4}f_1\tilde{f}_1 & \tilde{m}_1^2+\frac{g^2v^2}{4}\tilde{f}_1\tilde{f}_1 & -\frac{g^2v^2}{4}\tilde{f}_1f_2 & \frac{g^2v^2}{4}\tilde{f}_1\tilde{f}_2 &  \\\frac{g^2v^2}{4}f_0f_2 & \frac{g^2v^2}{4}f_1f_2 & -\frac{g^2v^2}{4}\tilde{f}_1f_2 & m_2^2+\frac{g^2v^2}{4}f_2f_2 & -\frac{g^2v^2}{4}f_2\tilde{f}_2 &  \\-\frac{g^2v^2}{4}f_0\tilde{f}_2 & -\frac{g^2v^2}{4}f_1\tilde{f}_2 & \frac{g^2v^2}{4}\tilde{f}_1\tilde{f}_2 & -\frac{g^2v^2}{4}f_2\tilde{f}_2 & \tilde{m}_2^2+\frac{g^2v^2}{4}\tilde{f}_2\tilde{f}_2 &  \\\vdots & \vdots &  &  &  & \ddots\end{array}\right)
\end{equation}
\begin{equation}\label{Mneutral}
\mathcal{M}_{\rm{neutral}}^2=\left(\scriptsize\begin{array}{cccccc}\frac{g^2v^2}{4c^2}f_0f_0 & \frac{g^2v^2}{4c^2}f_0f_1 & -\frac{g^2v^2c^\prime}{4c}f_0\tilde{f}_1 & \frac{g^2v^2}{4c^2}f_0f_2 & -\frac{g^2v^2c^\prime}{4c}f_0\tilde{f}_2 & \cdots \\\frac{g^2v^2}{4c^2}f_0f_1 & m_1^2+\frac{g^2v^2}{4c^2}f_1f_1 & -\frac{g^2v^2c^\prime}{4c}f_1\tilde{f}_1 & \frac{g^2v^2}{4c^2}f_1f_2 & -\frac{g^2v^2c^\prime}{4c}f_1\tilde{f}_2 & \cdots \\-\frac{g^2v^2c^\prime}{4c}f_0\tilde{f}_1 & -\frac{g^2v^2c^\prime}{4c}f_1\tilde{f}_1 & \tilde{m}_1^2+\frac{g^2v^2c^{\prime\,2}}{4}\tilde{f}_1\tilde{f}_1 & -\frac{g^2v^2c^\prime}{4c}\tilde{f}_1f_2 & \frac{g^2v^2c^{\prime\,2}}{4}\tilde{f}_1\tilde{f}_2 &  \\\frac{g^2v^2}{4c^2}f_0f_2 & \frac{g^2v^2}{4c^2}f_1f_2 & -\frac{g^2v^2c^\prime}{4c}\tilde{f}_1f_2 & m_2^2+\frac{g^2v^2}{4c^2}f_2f_2 & -\frac{g^2v^2c^\prime}{4c}f_2\tilde{f}_2 &  \\-\frac{g^2v^2c^\prime}{4c}f_0\tilde{f}_2 & -\frac{g^2v^2c^\prime}{4c}f_1\tilde{f}_2 & \frac{g^2v^2c^{\prime\,2}}{4}\tilde{f}_1\tilde{f}_2 & -\frac{g^2v^2c^\prime}{4c}f_2\tilde{f}_2 & \tilde{m}_1^2+\frac{g^2v^2c^{\prime\,2}}{4}\tilde{f}_1\tilde{f}_1 &  \\\vdots & \vdots &  &  &  & \ddots\end{array}\right).
\end{equation}   
Where here $\tilde{f}_n$ is the solutions of (\ref{gaugeeqn}) with $(-+)$ BC's and eigenvalues $\tilde{m}_n^2$. For ease of notation we have set $f_n=f_n\Theta_n$. Once again if the Higgs is localised on a codimension one brane then only the $l_i=0$ modes will contribute to the mass of the gauge zero modes. If on the other hand the Higgs is localised to a strict 3 brane then the $l_i\neq 0$ modes will not decouple and the components of the above mass matrices would become block matrices running over $f_nf_m\Theta_n(l_i=a)\Theta_m(l_i=b)$. Once again this would lead to uncomputable EW constraints.

As in the previous appendix one can compute the correction to the gauge zero mode masses by diagonalising the above matrices
\begin{eqnarray*}
\hat{M}^2_W\approx m_w^2\left (1-\sum_n\left [\frac{m_w^2F_n^2}{m_n^2-m_w^2(F_n^2-1)}+\frac{m_w^2\tilde{F}_n^2}{\tilde{m}_n^2+m_w^2(\tilde{F}_n^2-1)}\right ]+\mathcal{O}(m_n^{-4})\right )\\
\hat{M}^2_Z\approx \frac{m_w^2}{c^2}\left (1-\sum_n\left [\frac{\frac{m_w^2}{c^2}F_n^2}{m_n^2-\frac{m_w^2}{c^2}(F_n^2-1)}+\frac{c^{\prime\;2}m_w^2\tilde{F}_n^2}{\tilde{m}_n^2+m_w^2(c^{\prime\;2}\tilde{F}_n^2-c^{-2})}\right ]+\mathcal{O}(m_n^{-4})\right )
\end{eqnarray*}
Bearing in mind the relations $1-\frac{1}{c^2}=-s^{\prime\;2}$ and $1-c^{\prime\;2}=s^{\prime\;2}$ then one can see that the corrections to the W and Z masses will partially cancel and hence the T parameter will be suppressed. In fact this cancellation continues to the next order of the expansion as well,
\begin{eqnarray*}
\triangle M_Z^2-\triangle M_W^2\approx\sum_{n=1}\Bigg [\frac{\frac{m_w^2}{c^2}F_n^2}{m_n^2+\frac{m_w^2}{c^2}(F_n^2-1)}-\frac{m_w^2F_n^2}{m_n^2+m_w^2(F_n^2-1)}+\frac{c^{\prime\,2}m_w^2\tilde{F}_n^2}{\tilde{m}_n^2+m_w^2(c^{\prime\,2}\tilde{F}_n^2-\frac{1}{c^2})}\\
\qquad\qquad\qquad\qquad-\frac{m_w^2\tilde{F}_n^2}{\tilde{m}_n^2+m_w^2(\tilde{F}_n^2-1)}\Bigg ]+\sum_{n=1}\sum_{m\neq n}\Bigg [\frac{\left (\frac{1}{c^4}-1\right )m_w^4F_n^2F_m^2}{m_n^2m_m^2}+\frac{(c^{\prime\,4}-1)m_w^4\tilde{F}_n^2\tilde{F}_m^2}{\tilde{m}_n^2\tilde{m}_m^2}\\
\qquad\qquad\qquad\qquad\qquad\qquad\qquad\qquad+\frac{\left (\frac{c^{\prime\,2}}{c^2}-1\right )m_w^4F_n^2\tilde{F}_m^2}{m_n^2\tilde{m}_m^2}+\frac{\left (\frac{c^{\prime\,2}}{c^2}-1\right )m_w^4F_m^2\tilde{F}_n^2}{m_m^2\tilde{m}_n^2}\Bigg ]+\mathcal{O}(m_n^{-6}),
\end{eqnarray*}
where now the last four terms cancel using the relations $\frac{1}{c^4}-1=2s^{\prime\,2}+s^{\prime\,4}$, $c^{\prime\,4}-1=-2s^{\prime\,2}+s^{\prime\,4}$ and $\frac{c^{\prime\,2}}{c^2}-1=-s^{\prime\,4}$. Here we have used that the eigenvalues of a $N\times N$ symmetric matrix  with diagonal entries $A_n$ and off diagonal entries $B_{mn}$ are approximately 
\begin{displaymath}
\lambda_n\,\approx A_n-\sum_{i\neq n}^N\frac{B_{ni}^2}{A_i-A_n}+\sum_{i\neq n}^N\sum_{j\neq i}^N\frac{B_{ni}B_{nj}B_{ij}}{(A_i-A_n)(A_j-A_n)}+\mathcal{O}(A^{-3}).
\end{displaymath}
Even if these cancellations occur at every order in the expansion, as one would expect, it is important to note that in any realistic scenario the $T$ parameter is never exactly zero. The reason for this is simply due to the fact that we don't observe the additional $SU(2)$ force and hence it is always necessary to break the custodial symmetry. Here this breaking is done by imposing different boundary conditions on the $SU(2)_R$ fields than on the $SU(2)_L$ fields which in turn causes $\tilde{F}_n$ and $\tilde{m}_n$ to be different from $F_n$ and $m_n$.  In the RS model this difference is small, however it interesting to note that this difference increases when $\alpha <0.6$, see figure \ref{gaugemassesfig}. That is to say the shifting of the mass eigenvalues (i.e. the shifting of the root of the Bessel functions) enhances the extent to which the custodial symmetry is broken. Also the gauge fermion couplings are not protected by the custodial symmetry. In the gauge eigenstate, the gauge fermion coupling arises from the covariant derivative in $i\bar{\psi}\gamma^\mu D_\mu\psi$.
 \begin{eqnarray*}
D_\mu=\partial_\mu+\sum_{n}\bigg(-igf_\psi^{(n)}(T_L^+W_\mu^{+(n)}+T_L^-W_\mu^{-(n)})-ig\tilde{f}_\psi^{(n)}(T_R^+\tilde{W}_\mu^{+(n)}+T_R^-\tilde{W}_\mu^{-(n)}) \nonumber\\
-igsQf_\psi^{(n)}A_\mu^{(n)}-i\frac{g}{c}(T_L^3-s^2Q)f_\psi^{(n)}Z_\mu^{(n)}-i\tilde{f}_\psi^{(n)}(gc^{\prime}T_R^3-gs^{\prime}Q_X)\tilde{Z}_\mu^{(n)}\bigg ).
\end{eqnarray*}
Where $T_{L,R}^a$, and $Q_X$ are the charges under $SU(2)_{L,R}$ and $U(1)_X$, while $Q=T_L^3+T_R^3+Q_X$ and $T_{L,R}^\pm=(T_{L,R}^1\pm i T_{L,R}^2)$. Once again if one assumes all the fermions are located at the same point and one notes that the SM fermions must carry charge $T_R=0$. As before, if one rotates into the mass eigenstate with a suitable unitary matrix, then the corrections to the gauge fermion couplings can be absorbed by the field redefinitions 
\begin{eqnarray*}
W_\mu^{(0)}\rightarrow \left (1+\sum_n\frac{m_w^2F_nF_\Psi^{(n)}}{m_n^2+m_w^2(F_n^2-1)}+\mathcal{O}(m_n^{-4})\right )W_\mu^{(0)}\\
Z_\mu^{(0)}\rightarrow \left (1+\sum_n\frac{\frac{m_w^2}{c^2}F_nF_\Psi^{(n)}}{m_n^2+\frac{m_w^2}{c^2}(F_n^2-1)}+\mathcal{O}(m_n^{-4})\right )Z_\mu^{(0)}. 
\end{eqnarray*}

As before this leads to
\begin{eqnarray*}
\Pi_{11}(0)\approx\frac{v^2}{4}\sum_n\;\frac{m_w^2(2F_nF_\psi^{(n)}-F_n^2)}{m_n^2+m_w^2(F_n^2-1)}-\frac{m_w^2\tilde{F}_n^2}{\tilde{m}_n^2+m_w^2(\tilde{F}_n^2-1)}\\
\Pi_{33}(0)\approx\frac{v^2}{4}\sum_n\;\frac{\frac{m_w^2}{c^2}(2F_nF_\psi^{(n)}-F_n^2)}{m_n^2+\frac{m_w^2}{c^2}(F_n^2-1)}-\frac{m_w^2c^{\prime\;2}\tilde{F}_n^2}{\tilde{m}_n^2+m_w^2(c^{\prime\;2}\tilde{F}_n^2-c^{-2})}\\
\Pi_{3Q}^{\prime}=\Pi_{\gamma Z}^{\prime}=0\\
\Pi_{11}^\prime\approx-\frac{v^2}{4}\sum_n\frac{2F_nF_\psi^{(n)}}{m_n^2+m_w^2(F_n^2-1)}\\
\Pi_{33}^\prime\approx-\frac{v^2}{4}\sum_n\frac{2F_nF_\psi^{(n)}}{m_n^2+\frac{m_w^2}{c^2}(F_n^2-1)}
\end{eqnarray*}
Hence, as found in \cite{ Casagrande:2008hr, Casagrande:2010si, Delgado:2007ne}, the tree level contribution to $S$ parameter is approximately the same for the two models
 \begin{eqnarray}
S\approx -\frac{4M_Z^2c^2s^2}{\alpha}\sum_n\frac{2F_nF_\psi^{(n)}}{m_n^2+\frac{m_w^2}{c^2}(F_n^2-1)}+\mathcal{O}(m_n^{-4})\nonumber\\
T\approx\frac{1}{\alpha}\sum_n\Bigg (\frac{m_w^2(2F_nF_\psi^{(n)}-F_n^2)}{m_n^2+m_w^2(F_n^2-1)}-\frac{m_w^2\tilde{F}_n^2}{\tilde{m}_n^2+m_w^2(\tilde{F}_n^2-1)}-\frac{\frac{m_w^2}{c^2}(2F_nF_\psi^{(n)}-F_n^2)}{m_n^2+\frac{m_w^2}{c^2}(F_n^2-1)}\nonumber\\+\frac{m_w^2c^{\prime\;2}\tilde{F}_n^2}{\tilde{m}_n^2+m_w^2(c^{\prime\;2}\tilde{F}_n^2-c^{-2})}+\mathcal{O}(m_n^{-4})\Bigg ).\label{eqn:STcust}
\end{eqnarray}
These expressions could of course be simplified by, for example, neglecting the correction to the masses of the KK gauge modes from the Higgs, which would typically contribute as $\sim m_w^2/m_n^4$.  Alternatively one can again compare directly with observables, where once again the tightest constraint comes from the weak mixing angle,    
\begin{displaymath}
s_Z^2=\frac{\pi\alpha}{\sqrt{2}G_f}\frac{f_\psi^{(m)}(\mathcal{M}_{\rm{charged}}^2)^{-1}_{nm}f_\psi^{(n)}}{(f_\psi^{(0)})^2}.
\end{displaymath}
Repeating the method of the previous appendix, but now with the enlarged mass matrix, one finds that
\begin{equation}
\label{SZapp}
s_Z^2\approx s_p^2\left (1+\frac{c_p^2}{c_p^2-s_p^2}\sum_{n=1}\left [\frac{m_w^2F_\psi^{(n)\,2}}{m_n^2}-\frac{2m_w^2F_nF_\psi^{(n)}}{m_n^2}+s^{\prime\, 2}m_w^2\left (\frac{\tilde{F}_n^2}{\tilde{m}_n^2}-\frac{F_n^2}{m_n^2}\right )+\mathcal{O}(m_n^{-4})\right ]\right ).
\end{equation}

\bibliographystyle{JHEP}

\bibliography{Bibliography}

\end{document}